\documentclass[aps,twocolumn,nofootinbib,superscriptaddress,longbibliography]{revtex4-2}

\usepackage{amsmath}
\usepackage{amssymb}
\usepackage{epsfig,color}
\usepackage{graphicx}
\usepackage{dcolumn}
\usepackage{bm}
\usepackage{multirow}
\usepackage{bbm}
\usepackage{enumerate}
\usepackage{graphicx}
\usepackage{verbatim}
\usepackage{hyperref}
\usepackage{float}
\hypersetup{colorlinks=true,linkcolor=blue,anchorcolor=blue,citecolor=blue,filecolor=blue,urlcolor=blue,bookmarksnumbered=true,pdfview=FitB}
\usepackage[utf8]{inputenc}
\usepackage[english]{babel}
\usepackage[maxfloats=256]{morefloats}
\usepackage{mathptmx, newtxtext, newtxmath, xspace}
\newcommand{\ve}{\varepsilon}

\newcommand{\vf}{\varv_{\mathrm{F}}}
\newcommand{\kf}{k_{\mathrm{F}}}
\newcommand{\Ef}{E_{\mathrm{F}}}

\newcommand{\bk}{{\bf k}}
\newcommand{\bp}{{\bf p}}

\newcommand{\bq}{{\bf q}}
\newcommand{\bE}{{\bf E}}

\newcommand{\nn}{\nonumber}
\newcommand{\beq}{\begin{equation}}
\newcommand{\eeq}{\end{equation}}
\newcommand{\bea}{\begin{eqnarray}}
\newcommand{\eea}{\end{eqnarray}}
\newcommand{\bse}{\begin{subequations}}
\newcommand{\ese}{\end{subequations}}
\newcommand{\bwt}{\begin{widetext}}
\newcommand{\ewt}{\end{widetext}}

\newcommand{\qv}{{\bf q}}
\newcommand{\kv}{{\bf k}}

\newcommand{\bv}{{\bf v}}

\newcommand{\I}{\mathrm{Im}}
\newcommand{\R}{\mathrm{Re}}

\newcommand{\bsu}{\begin{subequations}}
\newcommand{\esu}{\end{subequations}}

\newcommand{\br}{{\bf r}}

\newcommand{\bQ}{{\bf Q}}

\newcommand{\eq}{\eqref}
\newcommand{\Eq}{Eq.~\eqref}

\newcommand{\vd}{\varv_\mathrm{D}}

\newcommand{\er}{\eqref}

\newcommand{\op}{\omega_\mathrm{p}}
\newcommand{\Ep}{T_\mathrm{p}}

\newcommand{\bi}{\begin{itemize}}
\newcommand{\ei}{\end{itemize}}

\begin{document}
\title{
Optical conductivity and damping of plasmons due to electron-electron interaction}
\author{Prachi Sharma}
\affiliation{Department of Physics, University of Florida, Gainesville, FL 32611-8440, USA}
\affiliation{ School of Physics and Astronomy, University of Minnesota, Minneapolis, MN 55455, USA} 
\author{Alessandro Principi}
\affiliation{Department of Physics and Astronomy, University of Manchester, Oxford Road, M13 9PL Manchester, UK} 
\author{Giovanni Vignale}
\affiliation{The Institute for Functional Intelligent Materials (I-FIM),
National University of Singapore, 4 Science Drive 2, Singapore 117544}
\author{Dmitrii L. Maslov}
\email{maslov@ufl.edu}
\affiliation{Department of Physics, University of Florida, Gainesville, FL 32611-8440, USA}

\date{\today}

\begin{abstract}
We re-visit the issue 
of plasmon
damping due to electron-electron interaction.
The plasmon linewidth
can related to the imaginary part of the charge susceptibility or, equivalently, to the real part
of the optical conductivity, $\R\sigma(q,\omega)$. 
Approaching the problem first
via a standard semi-classical Boltzmann equation, we show that  $\R\sigma(q,\omega)$ of  two-dimensional (2D) electron gas scales as $q^2T^2/\omega^4$ for $\omega\ll T$, which agrees with the results of Refs.~\cite{principi:2013} and \cite{Sharma:2021}
but disagrees with that of  Ref.~\cite{mishchenko:2004}, according to which
$\R\sigma(q,\omega) \propto q^2T^2/\omega^2$. 
To resolve this disagreement, we re-derive $\R\sigma(q,\omega)$ using the original method of Ref.~\cite{mishchenko:2004} 
for an arbitrary ratio $\omega/T$ and show that, while the last term is, indeed, present, it is subleading to the  $q^2T^2/\omega^4$ term. We give a physical interpretation of both leading and subleading contributions in terms of the shear and bulk viscosities of an electron liquid, respectively. 
We also calculate $\R\sigma(q,\omega)$ for a three-dimensional (3D) electron gas and doped monolayer graphene. We find that, with all other parameters being equal, finite temperature has the strongest effect on the plasmon linewidth in graphene, where it scales as $T^4\ln T$ for $\omega\ll T$. 
\end{abstract}

\maketitle

\section{Introduction}
\label{sec:Intro}
Collective modes of a Fermi liquid (FL)  are the direct manifestation of its many-body nature. In a charged FL, the most well-studied  mode is the plasmon. The plasmon dispersion and linewidth contain important information about the many-body dynamics in electron systems
and is also a crucial parameter for plasmonic devices.
Traditionally, plasmons have been observed by electron energy-loss spectroscopy \cite{Roth:2014}. The interest to plasmon dynamics has recently intensified due to near-field optical spectroscopy of graphene-based devices \cite{Woessner:2015,Ni:2016,Basov:2017} and momentum-resolved electron energy-loss spectroscopy of HTC cuprates and related compounds \cite{Vig:2017,*Mitrano:2018,*Thornton:2023,*chen:2023}. On the theoretical side, within the random-phase approximation (RPA) and at $T=0$ a plasmon has an infinitely long lifetime as along as it stays outside  the particle-hole continuum
and thus cannot decay via the Landau-damping mechanism.\footnote{At finite temperature, particle-hole pairs ``leak out'' through the continuum boundary, but the spectral weight of the leakage, given by the imaginary part of the charge susceptibility, is exponentially small: $\I\chi_c(q,\omega)\propto \exp\left[-\left(m\omega^2/q^2-\Ef)/2T\right)\right]$, where $m$ is the electron mass and $\Ef$ is the Fermi energy (see Appendix \ref{sec:RPA_T}). Consequently, the plasmon lifetime within RPA is finite but exponentially long.} Beyond RPA, plasmons do decay into multiple particle-hole pairs. In terms of Feynman diagrams, these processes are accounted for by dressing the free particle-hole bubbles with interaction lines. Damping of plasmons in three dimensions (3D) was studied by Dubois and M. G. Kivelson as early as in 1969 \cite{dubois:1969}. Damping of  plasmons in two dimensions (2D) was studied in Refs.~\cite{Reizer:2000,mishchenko:2004,principi:2013,briskot:2015,narozhny:2017,Lucas:2018,Sharma:2021} both in the collisionless   and hydrodynamic regimes. However, the results of different papers for damping in the  collisionless regime \cite{Reizer:2000,mishchenko:2004,principi:2013,Lucas:2018,Sharma:2021} do not always agree with each other, although all of them are obtained under the same assumptions, the most important of which is weak coupling.
The goal of this communication is to finalize the result for the lifetime of plasmons due to electron-electron interaction, at least at weak coupling.  We will limit our attention to the collisionless regime, which occurs if the plasmon frequency is much higher than the rate of relaxation towards equilibrium, and consider the cases of 2D and 3D electron gases with parabolic dispersion, as well as of doped monolayer graphene. We will also consider only damping due to intraband excitations, although interband excitations need to be accounted for to explain real materials \cite{Paasch:1970,*Gibbons:1977}.

Formally, damping of plasmons is due to the fact that the imaginary part of the charge susceptibility, $\chi_c(\bq,\omega)$ is finite outside the particle-hole continuum. Thanks to the Einstein relation,
\bea 
\label{Einsteineq}
\mathrm{Re}\sigma(q,\omega) = \frac{e^2 \omega}{
 q^2} \mathrm{Im} \chi_{c}
 (q,\omega), 
\eea
the same condition can be reformulated in terms of the real part of the conductivity. To facilitate the comparison, we list below the results of different papers for $\mathrm{Re}\sigma(q,\omega)$.\footnote{Reference \cite{Reizer:2000} did not account for all relevant diagrams, which led to a violation of gauge invariance, and will be not be discussed further.}

Mishchenko, Reizer, and Glazman (MRG) \cite{mishchenko:2004} considered a Galilean-invariant 2D electron gas (2DEG), i.e., a 2D electron system with $\epsilon_\bk=k^2/2m$ dispersion. Using an original method to calculate the absorption rate of electromagnetic radiation by electrons, they obtained the following result for the conductivity at finite $q$, $\omega$, and $T$: 
\bea
\text{Re}\sigma(q,\omega)=\frac{e^2}{12\pi^2}\frac{q^2}{k_F^2}\frac{\omega^2+4\pi^2 T^2}{\omega^2}\ln\frac{\vf\kappa}{\max\{|\omega|,T\}},\label{MRG}
\eea
\newline 
where $k_F$ and $\vf$ are the Fermi momentum and velocity, respectively, and $\kappa=2me^2$ is the inverse radius of the screened Coulomb interaction in 2D, defined by
\bea
V_\bQ=\frac{2\pi e^2}{Q+\kappa}.\label{vsc}
\eea
 Equation \eqref{MRG} is derived under the following assumptions: 
 $\max\{\omega,T\}\ll 
 \vf\kappa\lesssim \Ef$
 and $q\ll\max\{\omega,T\}/\vf$. (We remind the reader that $\vf\kappa\ll \Ef$ at weak coupling.)
 The results of the current paper will also be obtained under the same assumptions. Please note that henceforth $T$ stands for the temperature of the electron system, which may be different from the lattice temperature; 
for example,  Ref.~\cite{Ni:2016}
reports the electron temperature of graphene under near-field pumping to be as high as  3200\,K.

The vanishing of $\text{Re}\sigma(q,\omega)$ in Eq.~\er{MRG} at $q=0$ reflects the fact that internal forces  in a Galilean-invariant system do not affect the motion of the center-of-mass and, therefore, the dissipative part of the conductivity must vanish at $q=0$ and finite $\omega$.
Such a constraint is no longer valid for graphene, which is not a Galilean-invariant system. 
The optical conductivity of graphene away from the charge neutrality was analyzed in 
Refs.~\cite{principi:2013} and 
\cite{Sharma:2021}, which identified the limiting form of the conductivity at $q=0$.   In addition, 
Refs.~\cite{principi:2013} and \cite{Sharma:2021} presented an $\mathcal{O}(q^2)$ term, which is necessary for determining the plasmon linewidth. The complete result, as derived in Ref.~\cite{Sharma:2021}, reads 
\bse
\bwt
\bea
\R\sigma(q,\omega)&=&\R\sigma_1(\omega)+\R\sigma_2(q,\omega), \label{sigma12}\\
\R \sigma_1(\omega)
&= & \frac{e^2}{240} \frac{\omega^2}{\Ef^2} \left(1 + 4 \pi^2\frac{ T^2}{\omega^2}\right) \left(3 + 8\pi^2\frac{ T^2}{\omega^2}\right) \ln  \frac{\varv_\mathrm{D} \kappa}{\mathrm{max}\{|\omega|,2\pi T\}},\label{sigmaD}\\ 
\R \sigma_2(q,\omega)&=& \frac{e^2}{24 \pi^2} \frac{q^2 \kappa^2}{m^{*2}\omega^2}\left(1 + 4 \pi^2\frac{ T^2}{\omega^2}\right) \ln \frac{k_F}{\kappa},\label{P}
\eea
\ewt
\ese
where $\R\sigma_1(\omega)$ and $\R\sigma_2(q,\omega)$ are the $\mathcal{O}(q^0)$ and $\mathcal{O}(q^2)$ contributions, respectively,  $\vd$ is the Dirac velocity, which plays the role of $\vf$ for the Dirac spectrum, and $m^*=\kf/\vd$ 
.\footnote{The earlier form of Eq.~\eqref{P}, derived in Ref.~\cite{principi:2013},  did not capture the logarithmic factor and also had a different dependence on the coupling constant. In addition, Ref.~\cite{principi:2013} considered only the $T=0$ case.} Note that although $\R\sigma_1$ is finite at $q=0$, it is suppressed by a factor of $(\omega/\Ef)^2$ compared to a regular FL contribution \cite{gurzhi:1959}. \footnote{Equations \eqref{sigma12}-\eqref{P} are valid in the isotropic approximation, which neglects trigonal warping of the Fermi surfaces. If trigonal
warping is taken into account, there is another contribution to the conductivity, which is of a regular FL type. This contribution will be discussed in Sec.~\ref{sec:damping_graphene}.}

It was further conjectured in Refs.~\cite{principi:2013} and \cite{Sharma:2021}  that the $\mathcal{O}(q^2)$ contribution should be the same for the Dirac and parabolic spectra, up to the re-definitions of the effective mass and inverse screening length, i.e., the optical conductivity of a 2DEG should read
\bea
\R \sigma(q,\omega)&=& \frac{e^2}{24 \pi^2} \frac{q^2 \kappa^2}{m^{2}\omega^2}\left(1 + 4 \pi^2\frac{ T^2}{\omega^2}\right) \ln \frac{k_F}{\kappa}.\label{2DEG}
\eea
If this is the case (and we will show explicitly that it is),  then there is a contradiction with the MRG result, Eq.~\eqref{MRG}. Indeed,
Eqs.~\eqref{MRG} and \eqref{2DEG} differ by a factor of $(\vf\kappa/\omega)^2\sim \op^2(\kappa)/\omega^2$, where $\op(q)$ is the plasmon dispersion. This implies that Eq.~\eqref{2DEG} is parametrically larger than Eq.~\eqref{MRG} for $\omega\ll \op(\kappa)$.

The difference between Eqs.~\eqref{MRG} and \eqref{2DEG} is not purely mathematical. In fact, the corresponding contributions arise from different physical processes and, on a general level, are related to the bulk and shear viscosities of the electron liquid, correspondingly.

The real part of the optical conductivity is related to the plasmon linewidth, $\Gamma(q)$, defined by the complex dispersion relation $\omega=\op(q)-i\Gamma(q)$. In 2D \cite{mishchenko:2004},
\bea
\Gamma(q)=\pi q\R\sigma(q,\omega)\big\vert_{\omega=\op(q)},\label{G2D}
\eea
In a 2DEG, $\op(q)=\vf\sqrt{\kappa q/2}$. Accordingly, Eqs.~\eqref{MRG} and \eqref{P} give quite different results for the plasmon linewidth. At $T=0$, for example,
\bea
\Gamma(q)\big\vert_{\text{Eq}.~\eqref{MRG}}/\Gamma(q)\big\vert_{\text{Eq}.~\eqref{P}}\sim q/\kappa,
\eea
which is smaller then unity for $q\ll\kappa$.

The approaches employed in previous works \cite{mishchenko:2004,principi:2013,Sharma:2021} involve quite complicated computations. We find it instructive  to start with a more straightforward approach,
namely, with a semi-classical Boltzmann equation, which is valid for $\omega\ll T$.
In this regime, Eqs.~\eqref{MRG} and \eqref{2DEG} reduce to
\bea
\text{Re}\sigma(q,\omega)=\frac{e^2}{3k_F^2}\frac{q^2T^2}{\omega^2}\ln\frac{\vf\kappa}{T}\label{MRG0}
\eea
and
\bea
\text{Re}\sigma(q,\omega)=\frac{e^2\kappa^2}{6m^2}\frac{q^2T^2}{\omega^4} 
\ln\frac{k_F}{\kappa},\label{P0}
\eea
respectively, and it should be fairly easy to see which one is correct. This exercise is the subject of Sec.~\ref{sec:BE}.

The rest of the paper is organized as follows.
Section~\ref{sec:MRG_method} gives a brief review of the MRG method. In Sec.~\ref{sec:2D}, we re-derive the result for the optical conductivity of a 2DEG using the MRG method and show that, in agreement with conjectures of Refs.~\cite{principi:2013} and \cite{Sharma:2021}, it is given by Eq.~\eqref{2DEG} rather than Eq.~\eqref{MRG}. We must emphasize that the error in Ref.~\cite{mishchenko:2004} is purely computational and reflects neither on the MRG method itself nor on the results of this reference for plasmon damping due to electron-phonon interaction.  Using the MRG method, we also calculate the optical conductivity of a 3D electron gas in Sec.~\ref{sec:3D} and supply the details of the derivation of Eqs.~\eqref{sigma12}-\eqref{P} for graphene. In Sec.~\ref{sec:PhysicalDiscussion}, we give a physical interpretation of our results 
in terms of the bulk and shear viscosities of an electron liquid. In Sec.~\ref{sec:linewidth}, we discuss the plasmon linewidth.
Section \ref{sec:disc} presents our conclusions.
\section{
Optical conductivity via the semi-classical Boltzmann equation}
\label{sec:BE}
In this section, we calculate the longitudinal conductivity of a 2D Galilean-invariant electron system, using the semi-classical Boltzmann equation (BE). 
Assuming the electric field of the form ${\bf E}={\bf E}_0 e^{i (\bq\cdot\br-\omega t)} 
$, the BE for the Fourier transform of the non-equilibrium part of the distribution function $\delta f(\bq,\omega;\bk)\equiv \delta f_\bk$ reads
\bea
-i(\omega-\bv_\bk\cdot\bq+i 0^+) \delta f_{\bk} -e ({\bf E}_0\cdot\bv_\bk) n'_{\bk} = I_{\text{ee}}[\delta f_{\bk} ], \nn \\
\eea
where $n_\bk=n_\mathrm{F}(\ve_\bk)$ is the Fermi function, $n'_\bk=\partial n_\mathrm{F}(\ve_\bk)/
\partial\ve_\bk$, $I_{\text{ee}}[\delta f_{\bk} ]$ is the electron-electron collision integral, and an infinitesimally small imaginary term $i0^+$ was added to ensure the retarded nature of the response. With a definition $\delta f_\bk=n_\bk(1-n_\bk)g_\bk=-Tn_\bk' g_\bk$, the equation for $g_\bk$ reads \cite{abrikosov:book}
\bwt
\bea
i(\omega-\bv_\bk\cdot\bq+i 0^+) g_\bk -\frac{e}{T} ({\bf E}_0\cdot\bv_\bk)&=&\frac{1}{Tn'_\bk}\int_{\bk'\bp\bp'} W_{\bk,\bp\to\bk'\bp'}(1-n_{\bk'})(1-n_{\bp'})n_{\bp}n_{\bk}\delta(\ve_\bk+\ve_\bp-\ve_{\bk'}-\ve_{\bp
'})\nn\\
&&\times\delta(\bk+\bp-\bk'-\bp')\left(g_{\bk}+g_{\bp}-g_{\bk'}-g_{\bp'}\right),\label{BE}
\eea
\ewt
where $W_{\bk,\bp\to\bk'\bp'}$ is the scattering probability and $\int_\bk$ is a shorthand for $\int d^2k/(2\pi)^2$ (and similarly for other momenta).  The overall scale of the collision integral is given by the relaxation rate due to electron-electron interactions at finite $T$, $1/\tau_{\text{ee}}(T)
$. The temperature is assumed to be low enough so that the condition $1/\tau_{\text{ee}}(T)\ll\omega$ is satisfied, yet $\omega\ll T$. In this case,  Eq.~\eqref{BE} can be  solved by subsequent iterations in the collision integral. To zeroth order, we neglect the collision integral and obtain  
\bea
g^{(0)}_\bk=\frac{1}{T} \frac{e({\bf E}_0 \cdot\bv_\bk)}{i(\omega-\bv_\bk\cdot\bq+i 0^+)}.\label{g0}
\eea
At the next step, we substitute $g_\bk=g^{(0)}_\bk+g^{(1)}_\bk$ back into Eq.~\eqref{BE} and neglect $g^{(1)}_\bk$ inside the collision integral, to obtain \bwt
\bea
g^{(1)}_\bk &=& \frac{e}{ n'_{\bk} T^2 (\omega-\bv_\bk\cdot\bq+i0^+)}\int_{\bk'\bp\bp'} W_{\bk,\bp\to\bk'\bp'}(1-n_{\bk'})(1-n_{\bp'})n_{\bp}n_{\bk}\delta(\ve_\bk+\ve_\bp-\ve_{\bk'}-\ve_{\bp'})\nn\\
&&\times \delta(\bk+\bp-\bk'-\bp') \left(\frac{\bv_\bk}{\omega - \bv_\bk\cdot\bq }+ \frac{\bv_\bp}{\omega - \bv_\bp\cdot\bq} - \frac{\bv_{\bk'}}{\omega - \bv_{\bk'}\cdot\bq} - \frac{\bv_{\bp'}}{\omega - \bv_{\bp'}\cdot\bq} \right)\cdot\bE_0.\label{g1}\eea
\ewt
The conductivity is read off from the electrical current ${\bf j}=-e\int_{\bk} \delta f_\bk \bv_\bk=Te\int_{\bk} n'_\bk g_\bk \bv_\bk$, as a coefficient of linear proportionality between ${\bf j}$ and ${\bf E}$. As we are interested in the longitudinal part of the conductivity, we choose $\bq||\bE$. Then the 
conductivity can be found as
\bea
\R\sigma(q,\omega)=\frac{1}{2}\left[\R\sigma_{xx}(q,\omega)+\R\sigma_{yy}(q,\omega)\right],
\eea
where it is understood that $\sigma_{\alpha\alpha}(\bq,\omega)$ denotes the conductivity calculated with both $\bE$ and $\bq$ being along the $\alpha$-axis.

The real part of the zeroth order conductivity, obtained from Eq.~\eqref{g0},  is non-zero only within the particle-hole continuum, i.e., for $|\omega|<\vf q$, and is not relevant here, while the first-order correction in Eq.~\eqref{g1} yields
\bwt
\bea
\R\sigma(q,\omega) &=&\frac{e^2}{8T} \int_{\bk\bp\bk'\bp'}
 W_{\bk,\bp\to\bk'\bp'}(1-n_{\bk'})(1-n_{\bp'})n_{\bp}n_{\bk}\delta(\ve_\bk+\ve_\bp-\ve_{\bk'}-\ve_{\bp
'})
 \delta(\bk+\bp-\bk'-\bp')\nn\\
 &&\times\left[\frac{\bv
 _\bk}{\omega - \bv_\bk\cdot\bq }+ \frac{\bv
 _\bp}{\omega - \bv_\bp\cdot\bq} - \frac{\bv
 _{\bk'}}{\omega - \bv_{\bk'}\cdot\bq} - \frac{\bv
 _{\bp'}}{\omega - \bv_{\bp'}\cdot\bq} \right]^2.\label{sigma1}
\eea 
\ewt
To arrive at the last result, we used the symmetry properties of  $W_{\bk,\bp\to\bk'\bp'}$ \cite{levinson:book,pal:2012b}. At $\bq=0$, the square bracket in second line of  Eq.~\eqref{sigma1} is reduced to $\left(\bv_\bk +\bv_\bp-\bv_{\bk'}-\bv_{\bp'}\right)^2/\omega^2$, which vanishes identically for the Galilean-invariant case, when $\bv_\bk=\bk/m$. A finite result for the conductivity is obtained by expanding Eq.~\eqref{sigma1} in $q$. To order $q^2$, we obtain
\bwt
\bea
\R\sigma(q,\omega) &=&   \frac{e^2 q^2}{8T\omega^4} \int_{\bk\bp\bk'\bp'} W_{\bk,\bp\to\bk'\bp'}(1-n_{\bk'})(1-n_{\bp'})n_{\bp}n_{\bk}\delta(\ve_\bk+\ve_\bp-\ve_{\bk'}-\ve_{\bp
'})\delta(\bk+\bp-\bk'-\bp')\nn\\
&&\times \left[\bv_\bk(\bv_\bk\cdot \hat\bq)+\bv_\bp(\bv_\bp\cdot \hat\bq)-\bv_{\bk'}(\bv_{\bk'}\cdot \hat\bq)-\bv_{\bp'}(\bv_{\bp'}\cdot \hat\bq)\right]^2,\label{sigma2}
\eea
\ewt
where $\hat\bq=\bq/q$. The general form of $\R\sigma(q,\omega)$ can be deduced already at this step. Indeed, Eq.~\eqref{sigma2} contains integrals over three independent energies (say, $\ve_\bk$, $\ve_\bp$, and $\ve_{\bk'}$), each of them contributing a factor of $T$ to the final result. Therefore,
\bea
\R\sigma(q,\omega)\propto \frac{q^2 T^2}{\omega^4},\label{sigma3}
\eea
which is consistent with Eq.~\eqref{P0}.

The result \eqref{sigma3} can be understood in the following way. A factor of $q^2$ follows immediately from the facts that $\R\sigma(q,\omega)$ must vanish at $q=0$ and be analytic in $q$ (at finite $\omega$ and $T$). The scaling $1/\omega^4$ follows from the fact that we need to iterate the BE once and expand the result in $\vf q/\omega$. Finally, the factor of $T^2$ is the expected FL scaling of the scattering rate.

The rest of the calculation proceeds assuming  $W_{\bk,\bp\to\bk'\bp'}$ is given the Born approximation for the screened Coulomb potential \eqref{vsc}, i.e., $W_{\bk,\bp\to\bk'\bp'}=8\pi V^2_{\bk-\bk'}$. 
After a straightforward calculation (see Appendix \ref{app:sigma}), we arrive at
\bea
\R\sigma(q,\omega)= \frac{e^2  \kappa^2}{6 m^2 } \frac{q^2T^2}{\omega^4} \ln \frac{k_F}{\kappa},
\label{sigma4}
\eea
which coincides with Eq.~\eqref{P0} rather than Eq.~\eqref{MRG0}.
Given also that the conductivity must satisfy the first-Matsubara-frequency rule \cite{chubukov:2012,maslov:2012,maslov:2017b}, i.e., $\sigma(\omega=\pm 2\pi i T,T)=0$, one can generalize the result in Ref.~\eqref{sigma4} for the case of arbitrary ratio of $\omega$ to $T$ as $\R\sigma(q,\omega)\propto q^2\left(\omega^2+4\pi^2 T^2\right)/\omega^4$, which is Eq.~\eqref{2DEG}. In the next section, we will see that this is, indeed, the correct result.

\section{Optical conductivity via the Mishchenko-Reizer-Glazman (MRG) method}
\label{sec:EoM}

In this  section, we resolve the disagreement between the results 
of Refs.~\cite{mishchenko:2004}, and \cite{principi:2013} and \cite{Sharma:2021}, and finalize the correct expression for the optical conductivity of a 2DEG. Using the MRG method, we will show that, in addition to the contribution found by MRG [Eq.~\eqref{MRG}], there is also another contribution given by Eq.~\eqref{2DEG}. For completeness, we will also derive the expressions for the  optical conductivities of a 3D electron gas and graphene in Secs.~\ref{sec:3D} and \ref{sec:graphene}, respectively.

\subsection{MRG method}
\label{sec:MRG_method}

In the MRG method, one calculates the rate at which electromagnetic radiation is absorbed by a system of interacting electrons.
The differential probability of an electron-electron collision in the presence of 
a photon is written 
via the Fermi golden rule as
\bea
dw_{s,s_z} &=& 2 \pi |\mathcal{L}_{s,s_z}|^2 \delta(\ve_\bp+\ve_\bk-\ve_{\bp'}-\ve_{\bk'}+\omega)\nn \\
&\times& \delta(\bp+\bk-\bk'-\bp'+\bq)\frac{d^
D
\!p' d^
D
\!k'}{(2\pi)^2}\label{dwprob},
\eea
where $\bp$, $\bk$ ($\bp{'}$, $\bk{'}$) are the initial (final) momenta of electrons, $s$ is the total spin of two electrons in the initial state, $s_z$ is the spin projection on the quantization axis, and
$ \mathcal{L}_{s,s_z}$ is the matrix element which depends on $s$ and, in general, on $s_z$.  To first order in the screened Coulomb interaction, the matrix elements for the singlet and triplet states are given by 
\bea
\mathcal{L}_{0,0}&=& e \phi_0 \left( \frac{V_{\bk-\bk'} + V_{\bp' -\bk}}{\ve_\bp - \ve_{\bp+\bq}+\omega} + \frac{V_{\bk-\bk'} + V_{\bp -\bk'}}{\ve_{\bp'} - \ve_{\bp'-\bq}-\omega}\right.\nn \\
 &+&\left. \frac{V_{\bp'-\bp} + V_{\bp -\bk'}}{\ve_\bk - \ve_{\bk+\bq}+\omega}+\frac{V_{\bp'-\bp} + V_{\bp' -\bk}}{\ve_{\bk'} - \ve_{\bk'-\bq}-\omega}\right) 
\label{L0}
\eea
and 
\bea
\mathcal{L}_{1,0}&=& \mathcal{L}_{1,\pm 1}=e \phi_0 \left( \frac{V_{\bk-\bk'} - V_{\bp' -\bk}}{\ve_\bp - \ve_{\bp+\bq}+\omega} + \frac{V_{\bk-\bk'} - V_{\bp -\bk'}}{\ve_{\bp'} - \ve_{\bp'-\bq}-\omega} \right.\nn \\ 
&+& \left. \frac{V_{\bp'-\bp} - V_{\bp -\bk'}}{\ve_\bk - \ve_{\bk+\bq}+\omega}+\frac{V_{\bp'-\bp} - V_{\bp' -\bk}}{\ve_{\bk'} - \ve_{\bk'-\bq}-\omega}\right),
\label{L1}
\eea
respectively, where $q\phi_0$ is the in-plane component of the electric field of an electromagnetic wave.

Next, one derives the total probability of absorption using ~\Eq{dwprob} with matrix elements from Eq.~\eqref{L0}, which is then used to calculate the dissipation rate. The latter is then related to the real part of the longitudinal conductivity, which is given by \cite{mishchenko:2004}
\bea
\text{Re}\sigma(q,\omega) &=& \frac{
(1- e^{-\omega/T})}{ 4 q^2\omega^3 
\phi_0
^2
} \int \frac{d^D\!p d^D\!k d^D\!p' d^D\!k' }{(2\pi)^{3D-1}} \nn \\
&\times & (|\mathcal{L}_{0,0}|^2+3|\mathcal{L}_{1,0}|^2 ) 
 n_\bk n_\bp (1-n_{\bp'})(1-n_{\bk'}) \nn \\ 
&\times &\delta(\ve_\bp +\ve_\bk-\ve_{\bp'} -\ve_{\bk'}+\omega) \delta(\bp+\bk-\bk'-\bp'+\bq).\nn\\\label{sigma}
\eea 

The matrix elements can be written as $\mathcal{L}_{0,0} =e\phi_0(\mathcal{A}+\mathcal{A}_\mathrm{ex})/\omega^2$, and $\mathcal{L}_{1,0} =  e\phi_0(\mathcal{A}-\mathcal{A}_\mathrm{ex})/\omega^2$, where 
\bea
\mathcal{A} &=&\omega^2\left[V_{\bk-\bk'}\left( \frac{1}{\ve_\bp - \ve_{\bp+\bq}+\omega} + \frac{1}{\ve_{\bp'} - \ve_{\bp'-\bq}-\omega}\right)\right. \nn  \\
 &+&\left. V_{\bp'-\bp}\left(\frac{1 }{\ve_\bk - \ve_{\bk+\bq}+\omega}+\frac{1}{\ve_{\bk'} - \ve_{\bk'-\bq}-\omega}\right)\right]
, \label{generalA}
\eea
and $\mathcal{A}_\mathrm{ex}$ is the exchange term obtained by interchanging $\bp'\leftrightarrow \bk'$ in Eq.~\eqref{generalA}.\footnote{To make the analysis applicable to an arbitrary electronic dispersion, we defined $\mathcal{A}$ and $\mathcal{A}_{\text{ex}}$ without a factor of $1/m$, compared to the original MRG's notations \cite{mishchenko:2004}.
}  From now on, we will neglect the exchange term $\mathcal{A}_\mathrm{ex}$, which contains the  
interaction potential at large momenta transfers
and is, therefore, small for a 
weakly-screened Coulomb interaction.  
It is also convenient to introduce the momentum and energy transfers via $\bQ=\bp-\bp'=\bk'-\bk-\bq$ and 
$\Omega=\ve_\bp-\ve_{\bp'}=\ve_{\bk'}-\ve_{\bk}-\omega$, respectively, upon which Eq.~\eqref{sigma} becomes
\bwt
\bea
\text{Re}\sigma(q,\omega) &=& \frac{ 
{e^2} 
(1- e^{-\omega/T})}{ (2\pi)^{3D-1}q^2\omega^3} \int d{^D}\!Q d^D\!p d^D\!k d\Omega\,  \mathcal{A}^2\,
n_\mathrm{F}(\ve_\bk) n_\mathrm{F}(\ve_\bp) \left[1-n_\mathrm{F}(\ve_\bp-\Omega)\right]\left[1-n_\mathrm{F}(\ve_{\bk}+\Omega+\omega)\right] \nn\\
&&\times \delta(\ve_\bp-\ve_{\bp-\bQ}-\Omega)\delta(\ve_\bk-\ve_{\bk+\bQ+\bq}+ \Omega+\omega).
\label{sigma_wAgeneral}
\eea 
\ewt
Interestingly, 
the last formula can be expressed as a convolution of two  
free-electron
response functions.
For example,
as shown in Appendix~\ref{app:convolution}, 
the dominant contribution to the conductivity of a Galilean-invariant system can be cast into the following form
\bwt
\bea
\text{Re}\sigma(q,\omega) =  b_D \frac{e^2 q^2}{\omega^
{5}m^2} \int \frac{d{^D}\!Q Q^2}{(2\pi)^D} \int_{-\infty}^\infty \frac{d\Omega}{\pi}\, 
V_\bQ^2
\big[ n_\mathrm{B}(\Omega) - n_\mathrm{B}(\Omega - \omega) \big] \I\chi_T(Q,\Omega - \omega)
\I \chi_c (Q,\Omega),
\,\label{sigma_Convolution}
\eea 
where $n_\mathrm{B}(z)$ is the Bose function, $b_D = 8/15$ for $D = 3$ and $b_D = 1/2$ for $D=2$, and where we have introduced the
imaginary parts of the density-density response function~\cite{giuliani}
 \begin{equation} \label{spectrum_density_fluctuations}
 \text{Im} \chi_c (Q,\nu)\equiv -2\pi \int \frac{d^Dk}{(2\pi)^D} \left[n_\mathrm{F}
 (\ve_{{\bf k} -{\bf Q}/2})-n_\mathrm{F}(\ve_{{\bf k} +{\bf Q}/2})\right]\delta({\bf k}\cdot{\bf Q}/m- \nu)
 \end{equation}
and of the transverse current-current response function
\begin{equation} \label{spectrum_transverse_fluctuations}
 \text{Im} \chi_T (Q,\nu)\equiv -2\pi \int \frac{d^Dk}{(2\pi)^D}
 \left[n_\mathrm{F}
 (\ve_{{\bf k} -{\bf Q}/2})-n_\mathrm{F}(\ve_{{\bf k} +{\bf Q}/2})\right]
 |{\bf k}\times{\bf \hat Q}/m |^2 \delta({\bf k}\cdot{\bf Q}/m -\nu).
\end{equation}
\ewt
(The factors of two in the equations above account for spin degeneracy.)
In the zero-temperature limit the $\Omega$-integral is restricted to the range $0<\Omega<\omega$ (for positive $\omega$).
Furthermore, because 
both response functions 
vanish linearly at low frequency, we see that the integral goes as $\omega^3$, and the final result for the conductivity is proportional to $q^2/\omega^2$, as it should.
Similar formulas for the subdominant contributions to the conductivity of Galilean-invariant systems are provided in Appendix~\ref{app:convolution}.
Equation \eqref{sigma_Convolution}
helps to elucidate the nature of 
excitations 
responsible for plasmon damping, as will be  discussed in  detail in Section \ref{sec:PhysicalDiscussion}.

We now proceed with applying Eq.~\eqref{sigma_wAgeneral} to specific cases.

\subsection{Two-dimensional electron gas}
\label{sec:2D}
First, we consider a 2D electron gas with a parabolic dispersion
$\ve_\bk =k^2/2m$. As are we interested in the limit of $\vf q/\omega\ll 1$, we expand $\mathcal{A}$ in Eq.~\eqref{generalA} in $1/\omega$ as
\bse
\bwt
\bea
\mathcal{A}&=& \mathcal{A}_1+\mathcal{A}_2,\label{totA}\\
\mathcal{A}_1&=&V_{\bk-\bk'}\left[(\bv_\bp-\bv_{\bp'}) \cdot \bq  +\frac{q^2}{m}\right]+V_{\bp-\bp'}\left[(\bv_\bk-\bv_{\bk'}) \cdot \bq  +\frac{q^2}{m} 
\right],\label{A1}\\
\mathcal{A}_2&=&\frac{1}{\omega}\left\{V_{\bk-\bk'}\left[(\bv_\bp\cdot \bq)^2 - (\bv_{\bp'} \cdot \bq)^2
\right]+
V_{\bp-\bp'}\left[
(\bv_\bk\cdot \bq)^2 - (\bv_{\bk'} \cdot \bq)^2\right]\right\}.\label{A2}
\eea
\ewt
\ese
The $\mathcal{A}_1$ term in the equation above is the one that was found in Ref.~\cite{mishchenko:2004}. However, as will be shown below, one also needs to keep the $\mathcal{A}_2$ term, despite the fact that it appears to be next order in $1/\omega$. [Note that we have already neglected $\mathcal{O}(q^4)$ terms in $\mathcal{A}_2$.]

Expanding the interaction potential as $V_{\bQ+\bq}=V_{\bQ}+\bq\cdot\boldsymbol{\nabla}_\bQ V_{\bQ}$ and retaining only up to $\mathcal{O}(q^2)$ terms, we re-write $\mathcal{A}_1$ in Eq.~\eqref{A1} as
\bea
\mathcal{A}_1=\frac{1}{m}\left[q^2V_\bQ+(\bQ\cdot\bq)(\bq\cdot\boldsymbol{\nabla}V_\bQ)\right].
\label{A1_2}\eea
The MRG result, Eq.~\eqref{MRG}, is reproduced by keeping the first term in the equation above. 
Indeed, each of the three energy integrations in Eq.~\eqref{sigma_wAgeneral} (over $\ve_\bk$, $\ve_\bp$, and $\Omega$) contribute a factor of $\omega$, thereby canceling out a factor of $1/\omega^3$. Next, each of the two delta-functions contributes a factor of $1/Q$ which, in 2D, leads to a logarithmic divergence in the integral over $Q$ at the lower limit. Cutting off this divergence at $Q\sim \omega/\vf$, we reproduce the structure of Eq.~\eqref{MRG}.  Since the second term in Eq.~\eqref{A1_2} contains an extra factor of $Q$, the resultant $Q$-integration does not lead to a logarithmic divergence, and is thus subleading in the leading logarithmic sense.

Now, we turn to the ``new'' (compared to MRG) term, $\mathcal{A}_2$ in Eq.~\eqref{A2}. Expanding this term to order $q^2$, we obtain  
\bea
\mathcal{A}_2 &=& \frac{2V_\bQ}{m^2\omega}(\bq\cdot\bQ)\left[\bq \cdot(\bp-\bk-\bQ)\right].
\label{A'}
\eea
As to be expected (and indeed shown to be the case in Appendix \ref{app:sigma_MRG_2DEG}), a typical value of $|\bk-\bp|\sim k_F\gg Q\gtrsim \kappa$. Then $\mathcal{A}_2$ can be estimated as $|\mathcal{A}_2|\sim q^2Qk_FV_Q/m^2\omega$. The resultant integral over $Q$ is convergent at $Q=0$ but needs to be cut off at $Q\sim k_F$ at the upper limit, upon which one reproduces the structure of Eq.~\eqref{2DEG}. (The cross-term, $\mathcal{A}_1\mathcal{A}_2$, vanishes to leading order upon angular integration.)

We pause here to emphasize a non-trivial structure of the expansion in $1/\omega$. Indeed, the expansion of the individual components in Eq.~\eqref{generalA} starts with the $1/\omega$ terms, which cancel each other, followed by the $\mathcal{O}(\bv_{\bk,\bp}\cdot\bq/\omega^2)$ terms, which are supposed to be the leading ones. However, the latter also almost cancel each other, and one needs to keep two $\mathcal{O}(q^2)$ terms: the $q^2/2m$ terms in Eq.~\eqref{A1} and the entire $\mathcal{A}_2$ term in Eq.~\eqref{A2}. For $\omega\ll \op(\kappa)$, the ``new''  term ($\mathcal{A}_2$) exceeds the ``old'' one. 

Deferring all further details to Appendix \ref{app:sigma_MRG_2DEG}, we present here the final result for the optical conductivity of a 2DEG with parabolic dispersion:
\bwt
\bea
\text{Re}\sigma(q,\omega) &=&\frac{e^2}{48 \pi^2} \frac{q^2}{  k_F^2}\left(1 + 4 \pi^2\frac{ T^2}{\omega^2}\right) \ln
\frac{\vf \kappa}{\mathrm{max}\{|\omega|,2\pi T\}}
+ \frac{e^2}{24 \pi^2} \frac{q^2 \kappa^2}{m^2 \omega^2} \left(1 + 4 \pi^2\frac{ T^2}{\omega^2}\right) \ln
\frac{k_F}{\kappa}
.\label{result2DEG}
\eea
\ewt
The second (``new'') term 
coincides with Eq.~\eqref{2DEG}, which confirms the conjecture made in Refs.~\cite{principi:2013} and \cite{Sharma:2021}, while the second one is subleading to the first one for $\omega\ll\op(\kappa)$.
\subsection{Three-dimensional electron gas}
\label{sec:3D}
For a 3D electron gas with parabolic dispersion, the form of $\mathcal{A}$ as is the same in Eqs.~\eqref{totA}-\eqref{A'}, but the integrals are different due to a change of the phase space. Deferring the details to Appendix \ref{app:3DEG}, we present here only the final result for the optical conductivity in the limit of $\omega\ll\vf\kappa$:
\bea
\text{Re}\sigma(q,\omega) &= & \frac{e^2\kappa 
}{720}  \frac{  q^2 \kappa^2}{m^2 \omega^2 } \left(1 + 4 \pi^2\frac{ T^2}{\omega^2}\right)
,
\label{result3DEG}
\eea
where 
the inverse screening length is given by $\kappa^2 = 8 \pi e^2 N_F$,  and $N_F = mk_F/2\pi^2$ is the density of states per spin.
\subsection{Doped monolayer graphene}
\label{sec:graphene}
The optical conductivity of graphene was calculated in Refs.~\cite{principi:2013} and \cite{Sharma:2021}, but the most complete result was given in Ref.~\cite{Sharma:2021} without a derivation. Here, we re-derive this result using the MRG method.

Without loss of generality, we assume that the Fermi energy is located in the upper Dirac cone. In the low-energy limit, i.e., for $\max\{\omega,T\}\ll \Ef$, one can neglect the presence of the lower Dirac cone.\footnote{As shown in Ref.~\cite{Goyal:2023}, the interaction between electrons in the upper and lower Dirac cones gives a subleading (in the leading-logarithm sense) contribution to the optical conductivity.} Also, for a long-range Coulomb interaction, one can neglect processes that lead to swapping of electrons between the $K$ and $K'$ valleys, as well as the exchange parts of both intra-and intervalley scattering amplitudes. For the same reason, the phase factors in the matrix elements between spinor states can be replaced by unities. With all these simplification, electrons in graphene can be described by the following low-energy Hamiltonian 
\bwt
 \bea
\label{heh}
H_{0} =\sum_{ \varsigma, \bk,s} (\vd k-\Ef) c^{\dagger}_{\varsigma, \bk, s} c^{\phantom{\dagger}}_{\varsigma, \bk, s}
+\frac{1}{2} \sum_{\bk,\bp,\bQ}\sum_{s,s',\varsigma,\varsigma'} V^{(0)}_\bQ c^\dagger_{\varsigma,\bk+\bQ ,s}c^\dagger_{\varsigma', \bp-\bQ, s'} c^{\phantom{\dagger}}_{\varsigma',\bp, s'} c^{\phantom{\dagger}}_{\varsigma,\bk, s},
\eea
\ewt
where $\varsigma=\pm 1$ labels the $K$ and $K'$ valleys, $\bk$ is the electron momentum measured from the center of the corresponding valley, $s=\pm 1$ is the  spin projection, and $V^{(0)}_\bQ=2\pi e^2/Q$ is the bare Coulomb potential. Within this approximation, the valley index plays the role of a conserved isospin. Therefore, the optical conductivity can be calculated by applying the MRG method to then case of spin-1/2 fermions occupying a single valley, i.e., using Eq.~\eqref{sigma_wAgeneral} and then multiplying the result by $2^2=4$.

In contrast to 2D and 3D electron gases
graphene is a non-Galilean--invariant system. Therefore, its optical conductivity is finite even at $q=0$. However, we will see that in order to determine the plasmon linewidth accurately, one needs to retain both $\mathcal{O}(1)$ and $\mathcal{O}(q^2)$ terms in the optical conductivity.
Recalling the factor of $1/q^2$ in \Eq{sigma_wAgeneral}, we see that $\mathcal{A}$ should have terms of $\mathcal{O}(q)$ and  $\mathcal{O}(q^2)$, which would give the $\mathcal{O}(1)$ and $\mathcal{O}(q^2)$ contributions to the conductivity, respectively.
As shown in Appendix \ref{app:graphene}, the result for $\mathcal{A}$ to required accuracy is given by
\bse
\bea
\mathcal{A}&=& \mathcal{A}_1+ \mathcal{A}_2,\\
 \mathcal{A}_1 &=&
- V_\bQ \frac{q \vd}{\Ef}\left[(\ve_\bp- \ve_{\bp-\bQ}) \cos\theta_{\bq\bp}  + (\ve_\bk- \ve_{\bk+\bQ}) \cos\theta_{\bq\bk}\right],
\nn\\
\\
 \mathcal{A}_2 &=&
V_\bQ \frac{q^2 \vd}{2\kf}\left[\frac{1}{2}\sin^2\theta_{\bp\bq}+ 2\frac{\vd Q}{\omega} 
\left(F_\bp -F_\bk\right)\right],
\label{A_graphene}
\eea
\ese
where $$F_\bk = \cos\theta_{\bq\bk}(\cos\theta_{\bq\bQ} -\cos\theta_{\bq\bk}\cos\theta_{\bk\bQ}),$$ $F_\bp$ is obtained by interchanging $\bk$ with $\bp$ in  $F_\bk$, and $\theta_{{\bf n}{\bf n}'}$ denotes the angle between the vectors ${\bf n}$ and ${\bf n}'$. 

The $q$-independent part of $\R\sigma$ is exactly the same as calculated in Ref.~\cite{Sharma:2021}, and we will give only a brief summary of the computational steps.
Once $\mathcal{A}_1^2$ is substituted into Eq.~\eqref{sigma_wAgeneral}, the differences of electronic dispersions in Eq.~\eqref{A_graphene} can be expressed via the frequencies $\omega$ and $\Omega$, using the constraints imposed by the delta-functions in Eq.~\eqref{sigma_wAgeneral}.   Suppose that $T=0$. Then, given that $\Omega\sim \omega$, we have $\mathcal{A}_1^2\propto \omega^2$. The three energy integrations (over $\ve_\bk$, $\ve_{\bp}$, and $ \Omega$) are effectively constrained to the interval $(0,\omega)$ by the Fermi functions and, therefore, contribute a factor of $\omega$ each. Altogether, it follows that $\R\sigma_1(\omega)\propto \omega^2\times\omega^3/\omega^3=\omega^2$. This is the leading contribution to the conductivity for $\omega\ll \Ef$. Therefore, one can neglect the frequencies in the delta-functions, upon which they impose purely geometric constraints on the angular variables; for $Q\ll \kf$, we simply have $\cos\theta_{\bp\bQ}=\cos\theta_{\bk\bQ}=0$. The $Q$-integration is same the as for a 2DEG (see Appendix \ref{app:sigma_MRG_2DEG}): given that each of the two delta-function gives a factor of  $1/Q$, the integral over $Q$ diverges logarithmically at the lower limit. Cutting this divergence off at $Q=|\omega|/\vd$, we reproduce the $T=0$ limit of Eq.~\eqref{sigmaD}. The reasoning for the opposite limit of $\omega\ll T$ is essentially the same, except for now
the dispersions in $\mathcal{A}_1^2$ give a factor of $T^2$, another $T^3$ comes from the energy integrations, and the factor of $1-e^{-\omega/T}$ in Eq.~\eqref{sigma_wAgeneral} is reduced to $\omega/T$.  Cutting off the $Q$-integration at $Q=T/\vd$, we obtain $\R\sigma_1(\omega)\propto (\omega/T)\times T^5\times \ln T/\omega^3=T^4\ln T/\omega^2$, which is the $\omega\ll T$ limit of Eq.~\eqref{sigmaD}. 
Details of calculating the $\mathcal{O}(q^2)$ can be found in Appendix \ref{app:graphene}, and the final result is given by Eqs.~\eqref{sigma12}-\eqref{P}.

\section{Physical
interpretation}
\label{sec:PhysicalDiscussion}
To
clarify the physical 
meaning of the results obtained in the previous sections,  
we  
start from the 
relation
between
the real part of the conductivity 
and the viscosities of the electron 
liquid \cite{Conti:1999}. This relation can be inferred from the equation of motion for the  current density 
and has the form
\beq
\label{GeneralizedEinstein}
\text{Re}\sigma(q,\omega)  = e^2\frac{n  q^2}{m\omega^2}\left[\nu_L+\left(2-\frac{2}{D}\right)\nu_T\right]\,,
\eeq
where $\nu_L=\zeta/nm$ and $\nu_T=\eta/nm$ are the  bulk and shear viscosities, respectively, (also known as longitudinal and transverse kinematic viscosities respectively) of the electron 
gas 
with number density $n$, evaluated
in the collisionless regime. (Not to be confused with the hydrodynamic viscosities, which are non-perturbative in the Coulomb interaction). 

Interestingly, Eq.~(\ref{GeneralizedEinstein}) can be viewed as an extension of the Einstein relation for the conductivity. The standard Einstein relation (for electrons in the presence of impurities) reads 
\beq\label{SE}
\sigma = e^2 N(0)\mathcal{D}\,,
\eeq
where $N(0) = 
\lim_{q \to 0}\chi_c(q,0)$  is the density of states at the Fermi level, $\chi_c(q,\omega)$ is the charge susceptibility,
and $\mathcal{D}$ is the 
diffusion coefficient.
Equation~(\ref{GeneralizedEinstein})
can be obtained by replacing $N(0)= 
\lim_{q \to 0}\chi_c(q,0)$ by $\lim_{q\to 0} \chi_c(q,\omega)=-nq^2/m\omega^2$ at
finite frequency
 (notice, however, the change of sign in front of $\chi_c$) and $\mathcal{D}$ by $ \nu_L+\left(2-2/D\right)\nu_T$, which is the diffusion coefficient of the momentum density.

The key question now is: what is the low-frequency behavior of the transport coefficients $\nu_L$ and $\nu_T$?  The answer is that $\nu_T$ tends to a finite value at $\omega\to 0$ (plus corrections of order $\omega^2$), while  $\nu_L$ vanishes at 
$\omega\to 0$ as $\omega^2$.  Thus we see that the terms proportional to $\omega^{-2}$ arise from the shear viscosity, while the $\omega$-independent terms arise from the $\omega^2$ contributions to the shear viscosity as well as the bulk viscosity. 
(We stress that this is valid in dimensions $D>1$, where a transverse viscosity can be defined.)

What is the physical reason for the difference? 
The two viscosity coefficients can  be expressed in terms of the stress-stress response function as follows: 
\beq\label{nuL1}
\nu_L(\omega)=  -\frac{1}{D^2}\frac{\text{Im}\langle\langle \hat P_{\mu\mu};\hat P_{\nu\nu}\rangle\rangle_\omega}{nm\omega}
\eeq
and
\beq\label{nuT1}
\nu_T(\omega)= -\frac{\text{Im}\langle\langle \hat P_{xy};\hat P_{xy}\rangle\rangle_\omega}{nm\omega},
\eeq
where
\beq \label{eq:Kubo_prod}
\langle\langle \hat A;\hat B \rangle\rangle_\omega \equiv
- i \int_0^\infty dt \langle [\hat A(t), \hat B(0)]\rangle e^{i(\omega + i 0^+) t},
\eeq
$[\cdot, \cdot]$ is the commutator of two operators, and  $\langle \cdot\rangle$ denotes the statistical average over the equilibrium density matrix. 
The time evolution of $\hat A$ is generated by the non-interacting Hamiltonian. 
The spectral density  of an observable $\hat O$ is obtained from Eq.~\eq{eq:Kubo_prod} by calculating $\text{Im}\langle\langle \hat O;\hat O \rangle\rangle_\omega$.
Furthermore, the stress-tensor operator is given by \cite{Tokatly:2005}
\beq
\hat P_{\mu\nu}=\hat T_{\mu\nu}+\hat W_{\mu\nu},
\eeq
where
\beq
\hat T_{\mu\nu}= \sum_{{\bf k}} \frac{k_\mu k_\nu}{m}\hat a^\dagger_{{\bf k}}\hat a_{{\bf k}}^{\phantom{dagger}},
\eeq
\beq
\hat W_{\mu\nu}= \frac{1}{2} \sum_{{\bf k},{\bf Q}} \left(\frac{Q_\mu Q_\nu}{Q}
\frac{dV_Q^{(0)}}{dQ}
+\delta_{\mu\nu}
V_Q^{(0)}
\right)\hat a^\dagger_{{\bf k}}\hat 
n_{{\bf Q}}\hat a_{{\bf k}-{\bf Q}}^{\phantom{dagger}},
\eeq
$V_Q^{(0)}$ is the bare Coulomb potential, and $\hat 
n_{\bf Q} = \sum_{{\bf p}}\hat a^\dagger_{{\bf p}-{\bf Q}}\hat a^{\phantom{dagger}}_{{\bf p}}$ is the number density 
operator.

Let us consider the bulk viscosity first.  The trace of the stress tensor is easily seen to be given by $\hat P_{\mu\mu}=\hat T+\hat H$, where $\hat T$ is the kinetic energy operator and $\hat H$ is the 
total Hamiltonian of the electron 
system, i.e., the sum of the kinetic  
and interaction 
parts. $\hat H$ is a constant of the motion 
and thus does not contribute to the response function.  

Then we are left with
\beq
\nu_L(\omega)=  -\frac{1}{D^2}\frac{\text{Im}\langle\langle \hat T;\hat T\rangle\rangle_\omega}{nm\omega}\,,
\eeq  
which can be rewritten in terms of the time derivatives of $\hat T$ as 
\beq\label{nuL}
\nu_L(\omega)=  -\frac{1}{D^2}\frac{\text{Im}\langle\langle \dot{\hat T};\dot{\hat T}\rangle\rangle_\omega}{nm\omega^3}\,,
\eeq
where we used 
that
\bea \label{eq:derivative_Kubo}
i\omega \langle\langle \hat A;\hat B\rangle\rangle_\omega &=& i \langle [\hat A, \hat B]\rangle - \langle\langle \dot{\hat A};\hat B\rangle\rangle_\omega
\nonumber\\
&=&
i \langle [\hat A, \hat B]\rangle + \langle\langle \hat A; \dot{\hat B}\rangle\rangle_\omega.
\eea
The second line of this equation follows the from time-translational invariance of the response function. Equation~\eq{nuL} is obtained by noting that the commutator 
of two Hermitian operators is anti-Hermitian, and therefore all its eigenvalues are imaginary. Thus, the term  $i \langle [\dot{\hat T}, \hat T]\rangle$, obtained by 
applying Eq.~\eq{eq:derivative_Kubo} 
to $\langle\langle \hat T;\hat T\rangle\rangle_\omega$ twice, is purely real and can only contribute to the real part of the response function.

Due to the Coulomb interaction, the kinetic energy operator depends on time, and its time derivative 
is given by
\beq
\dot{\hat T} = -i\sum_{{\bf Q}} 
V^{(0)}_Q
({\bf Q}\cdot \hat {\bf j}_{-{\bf Q}}
)\hat 
n_{{\bf Q}}.
\eeq
(Notice that $\dot{\hat T}$ is proportional to the scalar product of the Coulomb force density $\hat {\bf F}_{{\bf Q}}=-i{\bf Q} 
V^{(0)}_Q\hat 
n_{{\bf Q}}$ 
and the \emph{longitudinal} current density $\hat {\bf j}_{-{\bf Q}}$.)

In the limit of a large number of fermion flavors \cite{Principi:2014}
the spectral function 
$\text{Im}\langle\langle \dot{\hat T};\dot{\hat T}\rangle\rangle_\omega$ is the convolution of two electron-hole spectral functions,
associated with 
density fluctuations
longitudinal current density fluctuations  
The spectral function
of density fluctuations vanishes as $\omega$ at $\omega\to 0$,
while that
of longitudinal current-density fluctuations vanishes as $\omega^3$, as one can see from the well-known relation 
$\text{Im}\langle\langle {\bf Q}\cdot \hat {\bf j}_{{\bf Q}}; {\bf Q}\cdot \hat {\bf j}_{-{\bf Q}}\rangle\rangle_\omega =\omega^2 \text{Im}\chi_c(Q,\omega)$ \cite{giuliani}.
Therefore, at zero temperature we have
\beq
\text{Im}\langle\langle \dot{\hat T};\dot{\hat T}\rangle\rangle_\omega \sim \int_0^\omega d\Omega~ \Omega^3 (\omega-\Omega) \propto \omega^5\,.
\eeq 
Substituting the last result into Eq.~(\ref{nuL}) gives $\nu_L
\propto\omega^2$ as announced. The essential reason for this result is the scarcity of longitudinal electron-hole pair excitations at low frequency: their spectral density vanishes as $\omega^3$.

Let us now consider the shear viscosity, Eq.~(\ref{nuT1}).  Without loss of generality we can orient the $x$ asis along ${\bf Q}$ and the $y$ axis perpendicular to ${\bf Q}$.  It can be easily shown that the averages $\langle\langle \hat W_{xy};\hat W_{xy} \rangle\rangle_\omega$ and $\langle\langle \hat T_{xy};\hat W_{xy} \rangle\rangle_\omega$ involve only longitudinal current density fluctuations and, therefore, vanish as $\omega^2$ as before.  However the average $\langle\langle \hat T_{xy};\hat T_{xy} \rangle\rangle_\omega$ 
now involves transverse current density fluctuations. 
As before, we rewrite $\nu_T(\omega)$ as
\beq\label{nuT}
\nu_T(\omega)=  -\frac{1}{D^2}\frac{\text{Im} \langle\langle \dot{\hat T}_{xy};\dot{\hat T}_{xy}\rangle\rangle_\omega}{nm\omega^3}\,
\eeq
and note 
that
\beq
\dot{\hat T}_{xy} 
-i \sum_{{\bf Q}}Q+\dots,
V^{(0)}_Q\hat 
j
_{T,-{\bf Q}}\hat 
n_{{\bf Q}},
\eeq
where $\hat  
j
_{T,-{\bf Q}}$ is the component of the current density 
in the $y$ direction, i.e., perpendicular to ${\bf Q}$
and  
$\dots$ stand for
subleading terms that contain the longitudinal component of the current-density operator.
Now we see that the spectral function $\text{Im}\langle\langle \dot{\hat T}_{xy};\dot{\hat T}_{xy}\rangle\rangle_\omega$  is the convolution of  two electron-hole spectral functions, one associated with 
density fluctuations 
and another one associated with {\it transverse}  current density fluctuations.
The spectral density of transverse current density fluctuations vanishes as $\omega$ as opposed to $\omega^3$.  Therefore we now have
\beq
\text{Im}\langle\langle \dot{\hat T}_{xy};\dot{\hat T}_{xy}\rangle\rangle_\omega \sim \int_0^\omega d\Omega~ \Omega (\omega-\Omega) \propto \omega^3\,.
\eeq   
Substituting the last result into Eq.~(\ref{nuT}) gives a finite value of $\nu_T$ in the $\omega \to 0$ limit. 

\begin{figure}[ht]
\centering
\includegraphics[width=1.0\columnwidth]{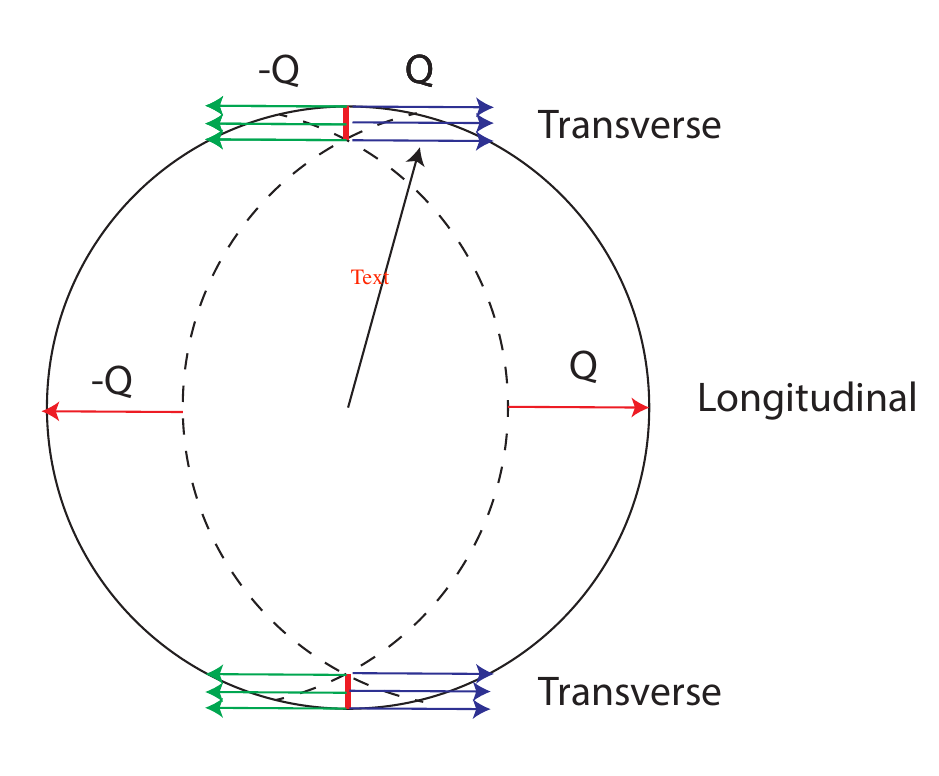}
\caption{
The 
horizontal arrows
represent the momenta of low-energy electron-hole pair excitations, 
${\bf Q}$ and $-{\bf Q}$.
Such excitations contribute to plasmon damping.  Due to kinematic constraints, the initial momenta of the excited electrons are confined to the crescent-shaped regions (external to the dashed lines), within the Fermi surface (solid circle).  The short red 
segments denote the widths of allowed regions of the initial momenta for excitations of a given (small) energy.   Notice that these excitations are essentially transverse, meaning that the average momentum of the initial and final state of the excited electron (black arrow)  is 
almost orthogonal to the momentum of the excitation, ${\bf Q}$.  The size of the integration region (red segments) vanishes linearly as the energy of the excitation tends to zero, consistent with the fact that the spectral 
function of density fluctuations
vanishes linearly at low frequency.  By contrast, longitudinal current excitations with the same wave vector ${\bf Q}$ (indicated by the red horizontal lines) have much higher energies: their contribution  to the low-frequency spectrum vanishes as the third power of the excitation energy.}
\label{fig1}
\end{figure}

The reason for the different behavior of the longitudinal and transverse current 
spectral functions is illustrated in Fig.~\ref{fig1}.

\section{Plasmon linewidth}
\label{sec:linewidth}
\subsection{Definitions}

We now derive the expressions for the plasmon linewidth, using the results $\R\sigma(q,\omega)$ obtained in previous sections. The linewidth, $\Gamma(q)$, is calculated 
at the plasmon pole and is given by  
\bea
\Gamma(q) = \pi q \R\sigma(q,\omega)\vert_{\omega=\omega_\mathrm{p}(q)} \label{damping2D}
\eea
in 2D and 
\bea
\Gamma(q) = 2 \pi  \R\sigma(q,\omega)\vert_{\omega=\omega_\mathrm{p}(q)}\label{damping3D}
\eea
in 3D.  
It is also customary to define a dimensionless inverse quality factor (IQF)
\bea
\gamma(q)=\frac{2\Gamma(q)}{\op(q)}.\label{IQF}
\eea

\subsection{Electron gas with parabolic dispersion}
\subsubsection{2D electron gas}
In a 2DEG, the plasmon frequency is $\omega_\mathrm{p}(q) = \vf \sqrt{\kappa q/2}$ with $\kappa = 2me^2$.  
Substituting the leading term in the optical conductivity [Eq.~\eqref{2DEG}] into Eq.~\eqref{damping2D}, we obtain the plasmon linewidth as 
\bea 
\Gamma(q) = \frac{e^2}{12\pi} \ln\frac{k_F}{\kappa}\frac{q^2 \kappa}{k_F^2} \left(1 + 4 \pi^2\frac{ T^2}{\omega_\mathrm{p}^2(q)}\right).\label{Gamma2DEG}
\eea
Defining $\Ep\equiv \vf\kappa\sim \op(\kappa)$, we see that $\Gamma(q)$
behaves as $qT^2$ for $q\ll\kappa (T/\Ep)^2$ and as $q^2$ for  $q\gg\kappa (T/\Ep)^2$. The corresponding IQF can be written in a dimensionless form as
\bea
\gamma(q)=\frac{2\sqrt{2}}{3\pi}\alpha^3_{\mathrm{e}}\ln\frac{1}{\alpha_{\mathrm{e}}}\bar q^{3/2}\left(1+\frac{8\pi^2 \bar T^2}{\bar q}\right),\label{gamma2D}
\eea
where $\alpha_\mathrm{e}=e^2/\vf$ is dimensionless coupling constant of the Coulomb interaction, $\bar q=q/\kappa$, and $\bar T=T/\Ep$.
\subsubsection{3D electron gas}
Neglecting plasmon dispersion, the plasmon frequency  of a For a 3D electron gas is given by $\omega_\mathrm{p} = \sqrt{4\pi n e^2/m}$,  where $n = k_F^3/3\pi^2$ is the number density. Substituting $\R\sigma$ from \Eq{result3DEG}  
into \Eq{damping3D}, we obtain 
\bea 
\Gamma(q) =\frac{\pi e^2\kappa}{360}\frac{q^2\kappa^2}{m^2\op^2}\left(1 + 4 \pi^2\frac{ T^2}{\omega_\mathrm{p}^2}\right),
\eea 
where $\kappa^2 = 8\pi e^2 N_F$ with $N_F = mk_F/2\pi^2$,
and
\bea
\gamma(q)=\frac{\sqrt{3}}{15}\alpha^2_{\mathrm{e}}\bar q^{2}\left(1+4\pi^2 \bar T^2\right),\label{gamma3D}
\eea
where the dimensionless parameters are the same as defined after Eq.~\eqref{gamma2D}. The asymptotic forms of $\Gamma(q)$ and $\gamma(q)$ are the same as for a 2DEG.
\subsection{Doped monolayer graphene}
\label{sec:damping_graphene}
So far, we discussed graphene in the isotropic approximation, which neglects trigonal warping of the Fermi surfaces around the $K$ and $K'$ points. The corresponding optical conductivity is given by the sum of $\R\sigma_1$ and $\R\sigma_2$ from Eqs.~\eqref{sigmaD} and \eqref{P}, respectively. Trigonal warping breaks the degeneracy of the $K$ and $K'$ valleys, and intervalley scattering is now allowed to contribute to the conductivity, even if there is no swapping of electrons between the valleys. The corresponding contribution to the conductivity was found in Ref.~\cite{Sharma:2021}:
\bea
\R\sigma_3(\omega)=\frac{29e^2}{48\pi^2}
\frac{e^2}{\vd}\ln\frac{\kf}{\kappa}(\kf a)^2\left(1+4\pi^2\frac{T^2}{\omega^2}\right),
\eea
where 
$a$ is the distance between two nearest carbon atoms. This contribution is of a regular FL type \cite{gurzhi:1959}, i.e., it is finite at $q=0$ and, in contrast to $\R\sigma_1$, is not suppressed due to partial Galilean invariance.
However, being a lattice effect, it becomes significant only at sufficiently high filling, when the product $\kf a$ is not too small.

Accordingly, the plasmon linewidth is written as the sum of three parts:
\bse
\bwt
\bea
\Gamma(q)&=&\Gamma_1(q)+\Gamma_2(q)+\Gamma_3(q),\\
\Gamma_1(q)&=&\frac{e^2}{
240\pi}\frac{q^2\kappa}{k^2_F} \left(1 + 4 \pi^2\frac{ T^2}{\omega_\mathrm{p}^2(q)}\right)\left(3 + 8 \pi^2\frac{ T^2}{\omega^2_\mathrm{p}(q)}\right)\ln
\frac{\vd\kappa}{\max\{\op(q),T\}},\label{Gamma1}\\
\Gamma_2(q)&=&\frac{e^2}{
24\pi}\frac{q^2\kappa}{k^2_F}\ln\frac{\kf}{\kappa} \left(1 + 4 \pi^2\frac{ T^2}{\omega_\mathrm{p}^2(q)}\right),\label{Gamma2}\\
\Gamma_3(q)&=&\frac{29e^2}{48\pi}
\frac{e^2q}{\vd}\ln\frac{\kf}{\kappa}(\kf a)^2\left(1+4\pi^2\frac{T^2}{\op^2(q)}\right),
\eea
\ewt
\ese
where  $\omega_\mathrm{p}(q) =\vd \sqrt{\kappa q}$ is the plasmon frequency and $\kappa = 4 e^2\kf/\vd$. For $q\ll\kappa\sqrt{T/\Ep}$, the linewidth is determined primarily by  $\Gamma_1(q)$ in Eq.~\eqref{Gamma1}, which 
is independent of $q$ and scales as $T^4\ln T$ in this limit. In the opposite limit of $q\gg\kappa\sqrt{T/\Ep}$, the contributions from $\Gamma_1$ and $\Gamma_2$  are of the same order-of-magnitude (although $\Gamma_2$ is numerically smaller) and both scale as $q^2$. On the other hand, the contribution from $\Gamma_3$ is much smaller than that from $\Gamma_1$ for any reasonable value of $\kf a$.

In terms of the dimensionless variables introduced after Eq.~\eqref{gamma2D}, the IQF for graphene is written as
\bwt
\bse
\bea
\gamma(q)&=&\gamma_1(q)+\gamma_2(q)+\gamma_3(q),\label{gammaG}\\
\gamma_1(q)&=&\frac{2\Gamma_1(q)}{\op(q)}=\frac{2\pi}{15} \bar q^{3/2}\alpha^3_\mathrm{g} \left(1+\frac{4\pi^2 \bar T^2}{\bar q}\right)\left(3+\frac{8\pi^2 \bar T^2}{\bar q}\right)\ln\frac{1}{
\max\{\bar q,\sqrt{\bar T}\}},\label{gamma1G} \\ 
\gamma_2(q)&=&\frac{2\Gamma_2(q)}{\op(q)}=\frac{
4}{3\pi}\bar q^{3/2}\alpha^3_\mathrm{g}\ln\frac{1}{\alpha_{\mathrm{g}}}\left(1+\frac{4\pi^2 \bar T^2}{\bar q}\right),\label{gamma2G}\\
\gamma_3(q)&=&\frac{2\Gamma_3(q)}{\op(q)}=\frac{29}{
24\pi} \bar q^{1/2}\alpha^2_{\mathrm{g}}\ln\frac{1}{\alpha_\mathrm{g}}(\kf a)^2\left(1+\frac{4\pi^2 \bar T^2}{\bar q}\right),\label{gamma3G}
\eea
\ese
\ewt
where $\alpha_\mathrm{g}=e^2/\vd$.

\subsection{Discussion}
\begin{figure}
    \includegraphics[width=1\linewidth]
    {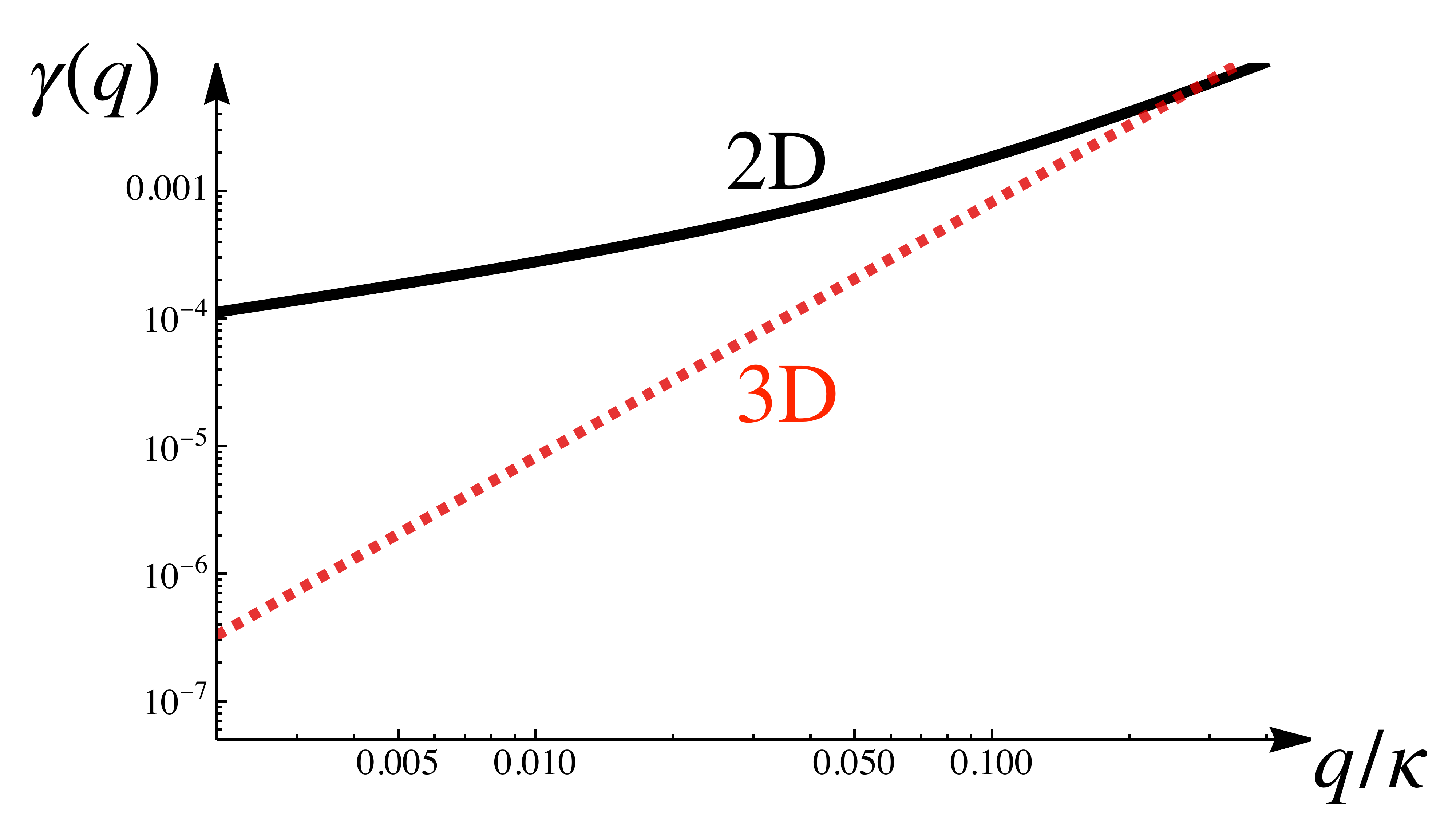}
    \caption{Inverse qualify factor, $\gamma(q)$ [Eq.~\eqref{IQF}], for 2D (dashed) and 3D (solid) electron gases as a function of $q/\kappa$, where $\kappa$ is the inverse screening radius. The electron temperature  $T=0.03\,\vf\kappa$, where $\vf$ is the Fermi velocity, and the coupling constant of the Coulomb interaction $\alpha=e^2/\vf=0.8$.}
    \label{fig:2D3D}
\end{figure}

The IQFs for 2D and 3D electrons gases are plotted as a function of $\bar q=q/\kappa$ in Fig.~\ref{fig:2D3D}. At other parameters being equal, damping is stronger in 2D (solid curve) than in 3D (dashed curve over a wide range of $q$. This is primarily due to the fact that finite temperature has little effect on damping is 3D. Indeed, the temperature dependence of $\gamma(q)$ in Eq.~\eqref{gamma3D} is weak as long as $\bar T\ll 1$ (or $T\ll\op$), and $\gamma(q)$ scales as $\bar q^{2}$ for any $\bar q$. 
On the contrary, $\gamma(q)$ in 2D [Eq.~\eqref{gamma2D}] decreases with $\bar q$ slower than in 3D, i.e., as $\bar q^{1/2}\bar T^2$ for $\bar q\ll \bar T^2$ (or
$q\ll \kappa\sqrt{T/\Ep}$). 
 As $\bar q$ approaches 1 from below, damping in 2D and 3D becomes comparable.

\begin{figure}
    \includegraphics[width=1\linewidth]
    {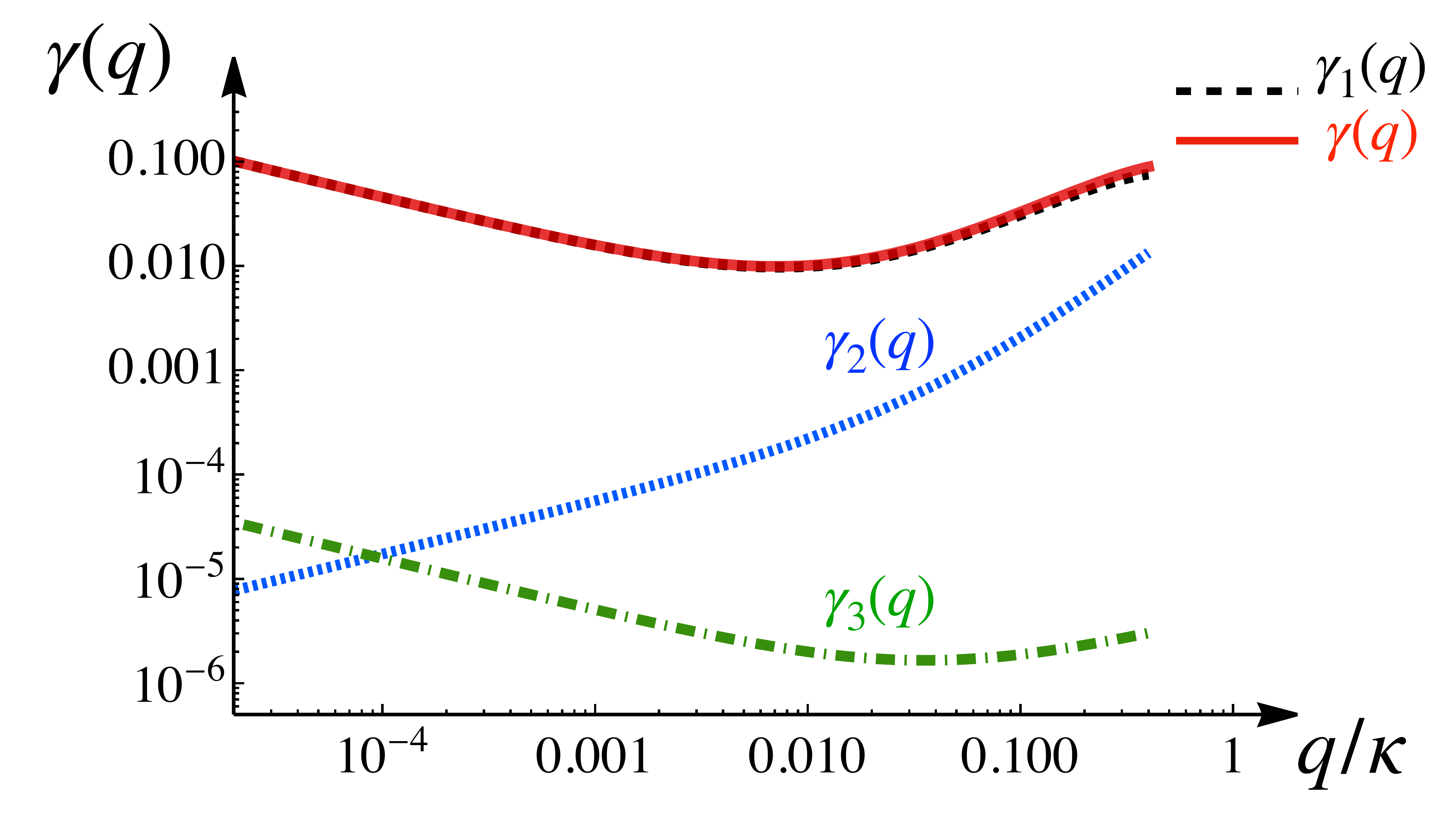}
    \caption{Inverse qualify factor, $\gamma(q)$ [Eq.~\eqref{IQF}], for doped monolayer graphene as a function of $q/\kappa$, where $\kappa$ is the inverse screening radius. $\gamma_1(q)$ (dashed), $\gamma_2$ (dotted), and  $\gamma_3$ (dash-dotted) are the partial components of total $\gamma(q)$ (solid), as defined by Eqs.~\eqref{gammaG}-\eqref{gamma3G}. The electron temperature  $T=0.03\,\vd\kappa$, where $\vd$ is the Dirac velocity, the filling factor $\kf a=0.1$, and the coupling constant of the Coulomb interaction $\alpha_{\mathrm{g}}=e^2/\vd=0.8$.}
    \label{fig:G}
\end{figure}

The partial components of $\gamma(q)$ as well as  total $\gamma(q)$ in graphene are plotted in Fig.~\ref{fig:G}. First, we note that $\gamma_1(q)$  dominates over the other two components for almost entire range of $q$. Next, we see that the effect of finite temperature on damping is graphene is even stronger than for a 2DEG: for $\bar q\ll 1$, $\gamma(q)\approx \gamma_1(q)$ scales as $\bar T^4/\sqrt{\bar q}$, i.e., the (relative) plasmon linewidth \emph{decreases} with increasing $\bar q$. In this regime, the usual numerical prefactor of $2\pi$, amplifying the effect of finite temperature on collective modes, plays a very important role, as it leads to an  enhancement of the linewidth by a factor of $2(2\pi)^4\approx 3120$. An apparent divergence of $\gamma(q)$ at $q\to 0$ signals a breakdown of the collisionless approximation and a crossover to the hydrodynamic regime. As $q$ increases, $\gamma(q)$ goes through a shallow minimum  at $\bar q=\left(4\pi^2(\sqrt{13}-2)/9\right)\bar T^2\approx 7.0 \bar T^2$ and then increases as $\bar q^{3/2}$. Comparing the vertical scales of Figs.~\ref{fig:2D3D} and \ref{fig:G}, we see that, at other parameters being equal, damping is, in general, stronger in graphene than in an electron gas with parabolic dispersion, even though the IQF for the latter is higher  order in the coupling constant (third vs second). This is, again, due to a higher sensitiviy of the linewidth of graphene to finite $T$. For $\bar q\ll \bar T^2$, the ratio $\gamma(q)\vert_{\mathrm{2DEG}}/\gamma(q)\vert_{\mathrm{graphene}}\sim \bar q/\bar T^2\ll 1$.

Damping of plasmons in graphene at finite $T$
was discussed by Lucas and Das Sarma \cite{Lucas:2018}, who conjectured that  $\Gamma(q)\propto q T^2$ in the collisionless regime. This result is consistent with Eq.~\eqref{Gamma2DEG} for a 2DEG with parabolic dispersion and with the subleading, $\Gamma_2(q)$ contribution to the linewidth in graphene [Eq.~\eqref{Gamma2}].
 However,  the leading contribution to the linewidth in graphene is given by $\Gamma_1(q)$ [Eq.~\eqref{Gamma1}], which is independent of $q$ and scales as $T^4\ln T$ for small $q\ll\kappa (T/\Ep)^2$).
 The discrepancy between our result and that of Ref.~\cite{Lucas:2018} is due to the assumption of Ref.~\cite{Lucas:2018} that the relevant relaxation rate in graphene scales in a canonical FL way, i.e., as $T^2$. However, whereas the single-particle relaxation rate in graphene indeed scales as $T^2$ (modulo a factor of $\ln T$), the current relaxation rate, defined as $\R\sigma=ne^2/m^*\omega^2\tau_{\mathrm{j}}(\omega,T)$ with $n$ being the carrier number density, scales as $1/\tau_{\mathrm{j}}\propto \max\{\omega^4\ln |\omega|,T^4\ln T\}$; see, e.g., Ref.~\cite{maslov:2017b} and references therein.
 This property is not unique to graphene but common for any 2D and 3D FL with an isotropic but non-parabolic spectrum (except for there is no $\ln T$ factor in 3D) \cite{Sharma:2021,Goyal:2023}.
\label{sec:dampingcoeff}
\section{Conclusions}
\label{sec:disc}
Prompted by some discrepancies between results of different theoretical papers, 
we re-visited the optical conductivity, $\R\sigma(q,\omega)$, of an electron system due to electron-electron interaction and a related issue of plasmon damping, focusing on the collisionless (as opposed to hydrodynamic) regime. Setting aside more sophisticated methods, we first showed that a  semi-classical Boltzmann equation for a two-dimensional electron gas (2DEG) with parabolic dispersion yields the optical conductivity which behaves as $q^2 T^2/\omega^4$ for $\vf q\ll \omega\ll T$. 
This behavior is consistent with the results of Refs.~\cite{principi:2013,Sharma:2021} but not with those of Ref.~\cite{mishchenko:2004}.   Next, we re-derived the full expression for $\R\sigma(q,\omega)$ of a 2DEG [Eq.~\eqref{2DEG}], using the original and elegant method of Ref.~\cite{mishchenko:2004}, and identified the reason for discrepancy between the results of Ref.~\cite{mishchenko:2004} and Refs.~\cite{principi:2013,Sharma:2021}.  We showed the results of Ref.~\cite{mishchenko:2004} and Refs.~\cite{principi:2013,Sharma:2021} correspond to physically distinct contributions, arising from the bulk and shear viscosities of an electron liquid. While the bulk viscosity vanishes at $\omega\to 0$, the shear one approaches a finite value. This explains why at sufficiently low frequency, the conductivity found in  Refs.~\cite{principi:2013,Sharma:2021} is larger than one found in Ref.~\cite{mishchenko:2004}.
 For completeness, we also calculated $\R\sigma(q,\omega)$ of a 3D electron gas and doped graphene, using the same method. The optical conductivity of a 3D electron gas [Eq.~\eqref{result3DEG}] is found to be similar to the 2D case: in both cases, $\R\sigma(q,\omega)$ vanishes at $q\to 0$ due to Galilean invariance. 
However, the case of doped graphene is different due to broken Galilean invariance. As a result, $\R\sigma(q,\omega)$ is finite at $q=0$ and scales as $\omega^2\ln|\omega|$ at $T=0$. This result was derived in our previous work \cite{Sharma:2021}, using the equations-of-motion method. Here, we also re-derived the $\mathcal{O}(q^2)$ term in the conductivity and found that it is exactly the same as in a 2D electron gas (up to a re-definition of an effective electron mass), in agreement with a conjecture of Refs.~\cite{principi:2013,Sharma:2021}.

Using our results for the optical conductivity, we analyzed the behavior of the plasmon linewidth, $\Gamma(q)$, in various systems. Given that the electron temperature in an experiment may be significantly higher than the lattice one, we paid special attention to the $T$-dependence of $\Gamma(q)$. We found that, with all parameters being equal, the effect of finite $T$ on plasmon damping is the strongest in graphene: $\Gamma(q)$ is independent of $q$ and scales as $T^4\ln T$ for $q\ll T^2/\vd^2\kappa$, where $\vd$ is the Dirac velocity and $\kappa$ is the inverse screening length. This scaling reflects the fact that the current relaxation rate in a non-Galilean-invariant but isotropic system behaves as $\max\{\omega^4,T^4\}$ (with an extra $\ln\max\{|\omega|,T\}$ factor in 2D) \cite{Sharma:2021,Goyal:2023}.
Plasmon linewidth in graphene can be measured via near-field spectroscopy \cite{Woessner:2015,Ni:2016}.  Also, a recent experiment has demonstrated that plasmons in a system with strong spin-orbit coupling can be observed directly by Raman spectroscopy \cite{sarkar:2023}. We hope that the results of our paper will be useful for the interpretation of these and future experiments.

\section{Acknowledgments}  
We are grateful to D. Basov, A. Chubukov, L. Glazman, E. Mishchenko, and M. Reizer for stimulating discussions. 
This work was supported by the US  National Science Foundation under Grant No. DMR-2224000 (D.L.M),
by the European Commission under EU Horizon 2020 MSCA-RISE-2019 programme (project 873028 HYDROTRONICS) and the 
Leverhulme Trust under grant RPG-2019-363 (A.P.), and by the Ministry of Education, Singapore, under its Research Centre of Excellence award to the Institute for Functional Intelligent Materials (I-FIM, project No. EDUNC-33-18-279-V12) (G.V.). During the completion of the manuscript, P.S. was supported by MRSEC through the DMR-2011401 grant. D. L. M. acknowledges the hospitality of the Kavli Institute for Theoretical Physics (KITP), Santa Barbara, supported by the National Science Foundation under Grants No. NSF PHY-1748958 and PHY-2309135.
\bibliography{plasmon_ref}
\appendix
\bwt
\section{Plasmon damping at finite temperature within the random-phase approximation}
\label{sec:RPA_T}
Within the RPA, the plasmon mode is the solution of the following equation:
\bea
1-V^{(0)}_\bq\Pi(q,\omega)=0,
\eea
where $V^{(0)}_\bq$ is the bare Coulomb potential and
\bea
\Pi(q,\omega)=2\int \frac{d^Dk}{(2\pi)^D} \frac{n_\mathrm{F}(\ve_\bk)-n_\mathrm{F}(\ve_{\bk+\bq})}{\omega-\ve_{\bk+\bq}+\ve_\bk+i 0^+}
\eea
is the polarization bubble (Lindhard function).
The plasmon linewidth is determined by the imaginary part of $\Pi(\bq,\omega)$.
For small $q$,
\beq
\I\Pi(q,\omega)=2\pi \omega \int \frac{d^Dk}{(2\pi)^D} n_\mathrm{F}'(\varepsilon_\bk) \delta(\omega -\qv\cdot\kv/m)\,.
\eeq
In two dimensions ($D=2$) and for a parabolic dispersion with mass $m$,
\bea
\I\Pi(q,\omega)&=&\omega\int_0^\infty\frac{dk k}{2\pi} n_\mathrm{F}'(\ve_\bk) \int_0^{2\pi}  d\phi\delta(\omega -qk \cos\phi/m)\nn\\
&=&\frac{m\omega}{\pi q}\int_{m|\omega|/q}^\infty dk~k \frac{n_\mathrm{F}'(\ve_\bk)}{\sqrt{k^2-\left(\frac{m\omega}{q}\right)^2}}.
\eea
Changing the integration variable from $k$ to $\epsilon=k^2/2m$, 
expressing $\epsilon$ in units of $\Ef$ as
$\epsilon=x\Ef$, and defining $\nu \equiv \omega/q\vf$, we obtain  
\bea
-\I\Pi(q,\omega)&=&\frac{m\nu}{\pi}\frac{\Ef}{T}\int_{\nu^2}^\infty \frac{dx}{\sqrt{x-\nu^2}}
 \frac{1}{4\cosh^2 \left[(x-1) \frac{\epsilon_\mathrm{F}}{2T}\right]}.
 \eea
 We are interested in the plasmon range, when $\nu\gg 1$, such that $x\geq \nu\gg 1$. In the degenerate case, we also have $\Ef\gg T$, thus the argument of $\cosh$ is always large, and $\cosh z\approx e^z/2$. The integral can be  then simplified as 
 \bea
 -\I\Pi(q,\omega)
 &
 \approx & \frac{m\nu}{\pi} \frac{\Ef}{T}e^{\Ef/T}\int_{\nu^2}^\infty \frac{dxe^{-x \Ef/T}}{\sqrt{x-\nu^2}}=
 \frac{m\nu}{\pi} \sqrt{\frac{\Ef}{T}}e^{-\frac{\Ef}{T}(\nu^2-1) }\int_{0}^\infty \frac{dy e^{-y}}{\sqrt{y}}\nn\\
 &=& \frac{m\nu}{\sqrt{\pi}} \sqrt{\frac{\Ef}{T}}e^{-\frac{\Ef}{T}(\nu^2-1) }=m\frac{\omega}{\vf q}\sqrt{\frac{\Ef}{\pi T}}e^{-\frac{\Ef}{T}\left(\frac{\omega^2}{\vf^2q^2}-1\right) }.
 \eea 
 (This result was presented in Ref.~\cite{mishchenko:2004} without a factor of $e^{\Ef/T}$.)

The plasmon linewidth is given by
\beq
\Gamma
(q) =- \frac{\omega_\mathrm{p}(q) V_q}{2}\Pi(q,\omega_\mathrm{p}(q)),
\eeq
where $\omega_\mathrm{p}(q)=\vf\sqrt{\kappa q/2}$ is the plasmon frequency, $\kappa=2me^2$ is the inverse Thomas-Fermi screening radius, and $V^{(0)}_\bq=2\pi e^2/q$.
According to Eq.~\eqref{IQF}, the inverse quality factor is given by
\beq
\gamma(q)=\frac{2\Gamma
(q)}{\omega_\mathrm{p}(q)} =  
\sqrt{\frac{\pi \Ef}{
2T}}\left(\frac{\kappa}{q}\right)^{3/2} e^{-\frac{\Ef}{T}\left(\frac{\kappa}{2q}-1\right)},
\eeq
which is exponentially small for $q\ll \kappa$ and $T\ll \Ef$.
\section{Derivation of Eq.~\eqref{sigma4}}
\label{app:sigma}
Substituting $W_{\bk,\bp\to\bk'\bp'}=8\pi V^2(|\bk-\bk')|)$ into \Eq{sigma2}, eliminating one of the four momenta in favor of the momentum transfer $\bQ=\bk'-\bk=\bp-\bp'$, and introducing the energy transfer via $\Omega=\ve_{\bk'}-\ve_{\bk}=\ve_{\bp}-\ve_{\bp'}$, we obtain  
\bea
\R\sigma(q,\omega) &=&   \frac{8\pi e^2 q^2 N^2_F}{8T\omega^4} \int\frac{d^2Q}{(2\pi)^2} \int d\Omega \int d\ve_\bk \int d\ve_\bp \int\frac{d\theta_{\bp\bQ}}{2\pi}\int\frac{d\theta_{\bk\bQ}}{2\pi} V^2_\bQ n_{\bk}n_{\bp}(1-n_{\bk+\bQ})(1-n_{\bp-\bQ}) \nn \\
&\times &\delta (\Omega +\ve_\bk -\ve_{\bk+\bQ}) \delta (\Omega -\ve_\bp +\ve_{\bp-\bQ})\left[\bv_\bk(\bv_\bk\cdot \hat\bq)+\bv_\bp(\bv_\bp\cdot \hat\bq)-\bv_{\bk+\bQ}(\bv_{\bk+\bQ}\cdot \hat\bq)-\bv_{\bp-\bQ}(\bv_{\bp-\bQ}\cdot \hat\bq)\right]^2 ,
\label{sigma2_app}
\eea
where 
$N_F= m/2\pi$ is the density of states, and $\theta_{{\bf a}{\bf b}}$ is the angle between vectors ${\bf a}$ and ${\bf b}$. Expanding the factor in the square brackets in \Eq{sigma2_app} in $Q/k_F$ to order $Q^2$ and projecting the electron momenta onto the Fermi surface, we obtain 
\bea
\left[\cdots \right]^2&=& \frac{k^2_FQ^2}{m^4}\left(\hat{\bQ} \cos\theta_{\bk\bq}+\hat{\bk} \cos\theta_{\bq\bQ}- \hat{\bQ}\cos\theta_{\bp\bq}-\hat{\bp} \cos\theta_{\bq\bQ} \right)^2.
\eea
Writing $\theta_{\bq\bp} = \theta_{\bq\bQ} + \theta_{\bQ\bp}$, $\theta_{\bq\bk}=\theta_{\bq\bQ} + \theta_{\bQ\bk}$ and integrating over $\theta_{\bq\bQ}$, we get 
\bea
\int_{0}^{2\pi}\frac{d\theta_{\bq\bQ}} {2\pi}\left[\cdots\right]^2&=&2\frac{k^2_FQ^2}{m^4}\left[3-\cos\left(\theta_{\bk\bQ}+\theta_{\bp\bQ}\right)\right]\sin^2\frac{\theta_{\bk\bQ}-\theta_{\bp\bQ}}{2}.\label{sqbrac}
\eea

To integrate over $\theta_{\bk\bQ}$ and $\theta_{\bp\bQ}$, we use the constraints imposed by energy conservation via the delta-functions in Eq.~\eqref{sigma2_app} With $\Omega=0$ and $\mathcal{O}(Q^2)$ terms discarded, these constraints are $\theta_{\bk\bQ}= \pm \pi/2$ and $\theta_{\bp\bQ}= \pm \pi/2$. The last term in Eq.~\eqref{sqbrac} ensures that only the combinations of $\theta_{\bk\bQ}=-\theta_{\bp\bQ}=\pm \pi/2$ give non-zero contributions. Summing up these contributions, we obtain
\bea
\int_{0}^{2\pi}
{d\theta_{\bk\bQ}}
\int_{0}^{2\pi}
{d\theta_{\bp\bQ}}
\left[\cdots\right]^2&=&\frac{8k^2_FQ^2}{m^4}.\label{int_angle}
\eea%
Next, the energy integration gives 
\bea
\int d\Omega \int d\ve_\bk \int d\ve_\bp \left[1-n_\mathrm{F}(\ve_{\bk}+\Omega)\right]\left[1-n_\mathrm{F}(\ve_{\bp}-\Omega)\right]n_\mathrm{F}(\ve_{\bp})n_\mathrm{F}(\ve_{\bk}) = \frac{2\pi^2 T^3}{3}. \label{int_freq}
\eea 
Substituting Eqs.~\eqref{int_angle} and \eqref{int_freq} back into Eq.~\eqref{sigma2_app}, and calculating the $Q$-integral to leading logarithmic accuracy with an upper limit cutoff at $Q\sim k_F$, we arrive at the final result
\bea
\R\sigma(q,\omega) &=&   
 \frac{ e^2 \kappa^2}{6 m^2 }  \frac{q^2T^2}{\omega^4}\int^{k_F}_0
 \frac{dQQ}{(Q+\kappa)^2}\approx \frac{ e^2 \kappa^2}{12 m^2 }  \frac{q^2T^2}{\omega^4}\ln \frac{k_F}{\kappa},
\eea
which is \Eq{sigma4} of the main text.

\section{Optical conductivity via MRG method}
\subsection{2D electron gas} 
\label{app:sigma_MRG_2DEG}
   
In this Appendix, we derive Eq.~\eqref{result2DEG} of the main text. We start with a 2D version of Eq.~\eqref{sigma_wAgeneral} with $\mathcal{A}$ given by the sum of the first term in Eq.~\eqref{A1_2} and the entire Eq.~\eqref{A'}:
\bea
\mathcal{A} &=&\frac{V_\bQ}{m}\left\{ q^2+2\frac{(\bq\cdot\bQ)\left[\bq \cdot(\bp-\bk-\bQ)\right]}{m\omega} \right\}.
\label{A'_app}
\eea 
Replacing the integration over $\bp (\bk)$ integral by that over $\ve_\bp (\ve_\bk)$ and angles $\theta_{\bp\bQ}(\theta_{\bk\bQ})$, and confining the energy integrations to the narrow vicinity of the Fermi energy, we obtain
\bea
\text{Re}\sigma(q,\omega) &=&\frac{e^2 (1- e^{-\omega/T})}{ (2\pi)^3\omega^3 q^2} N_F^2\int d^2Q \int^\infty_{-\infty}d \ve_\bp \int^\infty_{-\infty}d\ve_\bk  \int^\infty_{-\infty}d\Omega 
\int_{0}^{2\pi} d\theta_{\bp\bQ} \int_{0}^{2\pi} d\theta_{\bk\bQ} \mathcal{A}^2(\theta_{\bk\bQ},\theta_{\bp\bQ}) \nn \\
&\times& n_\mathrm{F}(\ve_\bp) n_\mathrm{F}(\ve_\bk) \left[1-n_\mathrm{F}(\ve_\bk -\Omega)\right]\left[1-n_\mathrm{F}(\ve_\bp+\Omega+\omega)\right] \delta\left(\bv_\bp\cdot\bQ- Q^2/2m +(\Omega+\omega)\right) \delta( \bv_\bk\cdot\bQ + Q^2/2m +\Omega), \nn \\ \label{s22}
\eea
where $N_F=m/2\pi$ is the density of states per spin orientation.
Writing $\theta_{\bq\bp} = \theta_{\bq\bQ} + \theta_{\bQ\bp}$ and $\theta_{\bq\bk}=\theta_{\bq\bQ} + \theta_{\bQ\bk}$, and folding the angular integrations down to the $(0,\pi)$ intervals, we get
\bea
\text{Re}\sigma(q,\omega) &=&\frac{e^2 (1- e^{-\omega/T}) 
}{ (2\pi)^3\omega^3 q^2} N_F^2\int^\infty_{-\infty}d \ve_\bk \int ^\infty_{-\infty}d \ve_\bp   \int^\infty_{-\infty} d\Omega \int d^2 Q  \nn \\ 
&\times &\int_{0}^{\pi} d\theta_{\bp\bQ} \int_{0}^{\pi} d\theta_{\bk\bQ} \left[\mathcal{A}^2(\theta_{\bk\bQ}, \theta_{\bp\bQ}) +\mathcal{A}^2(-\theta_{\bk\bQ},- \theta_{\bp\bQ}) + \mathcal{A}^2(-\theta_{\bk\bQ}, \theta_{\bp\bQ}) + \mathcal{A}^2(\theta_{\bk\bQ}, -\theta_{\bp\bQ})  \right]  \nn \\ 
&\times& n_\mathrm{F}(\ve_\bp) n_\mathrm{F}(\ve_\bk) (1-n_\mathrm{F}(\ve_\bk -\Omega))(1-n_\mathrm{F}(\ve_\bp+\Omega+\omega)) \delta(\bv_\bp\cdot\bQ+ Q^2/2m -(\Omega+\omega)) \delta( \bv_\bk\cdot\bQ - Q^2/2m - \Omega), \nn \\
 \label{s22_1}
\eea 
where  
\bea
\mathcal{A}(\theta_{\bk\bQ},\theta_{\bp,\bQ}) &=&\frac{ q^2 V_\bQ}{m} \left\{1 -2\frac{Q \cos \theta_{\bq\bQ}}{m\omega}{\left[\kf \cos(\theta_{\bq\bQ} + \theta_{\bQ\bp})-\kf\cos(\theta_{\bq\bQ} + \theta_{\bQ\bk}) + Q \cos \theta_{\bq\bQ}\right]}\right\},
\label{A}
\eea
is the projection of $\mathcal{A}$ onto the Fermi surface.

To perform the angular integrations  in \Eq{s22_1}, we use the constraints imposed by the delta functions in Eq.~\eqref{A}, i.e., 
\bea
\theta_{\bk\bQ} &=& \cos^{-1}\left[ \frac{\Omega}{\varv_{\text{F}} Q} + \frac{Q}{2\kf}\right],\nn\\ 
\theta_{\bp\bQ} &=& \cos^{-1}\left[ \frac{(\Omega+\omega)}{\vf Q}- \frac{Q}{2\kf}\right].\label{cos}
\eea
Simple trigonometry yields
 \bea
 \cos(\theta_{\bq\bQ} \pm \theta_{\bQ\bp})
 &=&\cos\theta_{\bq\bQ} \left(\frac{\Omega+\omega}{\vf Q}- \frac{Q} {2\kf}
 \right)\mp \sin\theta_{\bq\bQ} \sqrt{1- \left(\frac{ (\Omega+\omega)}{\vf Q}- \frac{Q} {2\kf} 
 	\right)^2}, \nn \\
 \cos(\theta_{\bq\bQ} \pm \theta_{\bQ\bk}) &=&\cos\theta_{\bq\bQ} \left(\frac{ \Omega}{v_k Q}+ \frac{Q} {2\kf}
 \right) \mp \sin\theta_{\bq\bQ} \sqrt{1- \left(\frac{ \Omega}{\vf Q}+ \frac{Q} {2\kf}
 	\right)^2}.
  \label{B6}
 \eea
Substituting Eq.~\eqref{B6} back into Eq.~\eqref{A}, we get 
\bse
\bea
\mathcal{A}(\theta_{\bk\bQ}, \theta_{\bp\bQ})  &=&\mathcal{A}(-\theta_{\bk\bQ}, -\theta_{\bp\bQ}) =\frac{q^2 V_\bQ}{m}\nn\\
&\times& \left\{ 1 -2 \cos^2 \theta_{\bq\bQ}  
- 2\frac{\kf Q}{m\omega}\cos \theta_{\bq\bQ}\sin \theta_{\bq\bQ} \left[\sqrt{1- \left(\frac{ \Omega+\omega}{\vf Q}- \frac{Q} {2\kf} 
	\right)^2}- \sqrt{1- \left(\frac{ \Omega}{\vf Q}+ \frac{Q} {2\kf}
	\right)^2}\right] \right\}, \label{Aa}\\
\mathcal{A}(\pm\theta_{\bk\bQ},\mp \theta_{\bp\bQ}) &=&\frac{q^2 V_\bQ}{m} \left\{ 1 -2 \cos^2 \theta_{\bq\bQ}  
\mp 2\frac{\kf Q}{m\omega}\cos \theta_{\bq\bQ}\sin \theta_{\bq\bQ} \left[ \sqrt{1- \left(\frac{ \Omega+\omega}{\vf Q}- \frac{Q} {2\kf} 
	\right)^2}+  \sqrt{1- \left(\frac{ \Omega}{\vf Q}+ \frac{Q} {2\kf}
	\right)^2}\right] \right\}. \nn \\
\label{As}
\eea
\ese
We anticipate (and will prove it later) typical values of the energy and momentum transfers to be  such that $\Omega\sim \omega\ll \vf Q$ and $Q\ll \kf$. If so, then the factor in square brackets in Eq.~\eqref{Aa} is small as
\bea
\sqrt{1- \left(\frac{ \Omega+\omega}{\vf Q}- \frac{Q} {2\kf} 
	\right)^2}- \sqrt{1- \left(\frac{\Omega}{\vf Q}+ \frac{Q} {2\kf}\right)^2}
 \approx -\frac{1}{2}\frac{\omega+2\Omega}{\vf Q}\left(\frac{\omega}{\vf Q}-\frac{Q}{k_F}\right)
\eea
and can be neglected.
Under the some conditions, the factor in square brackets in Eq.~\eqref{As} is almost equal to 2. With these simplifications, we have
\bea
\mathcal{A}(\theta_{\bk\bQ}, \theta_{\bp\bQ})&=&\mathcal{A}'(-\theta_{\bk\bQ},- \theta_{\bp\bQ})   \approx \frac{q^2 V_\bQ}{m}\left( 1 -2 \cos^2 \theta_{\bq\bQ} \right), \nn \\
\mathcal{A}(\pm \theta_{\bk\bQ},\mp \theta_{\bp\bQ})  &\approx& \frac{q^2 V_\bQ}{m} \left( 1 -2 \cos^2 \theta_{\bq\bQ} \mp 4\frac{\vf Q}{\omega} \cos \theta_{\bq\bQ}\sin \theta_{\bq\bQ}  \right).\label{19}
\eea  
The square brackets in Eq~(\ref{s22_1}) can now be written as 
\bea
\mathcal{A}^2(\theta_{\bk\bQ}, \theta_{\bp\bQ}) +\mathcal{A}^2(-\theta_{\bk\bQ},- \theta_{\bp\bQ}) + \mathcal{A}^2(-\theta_{\bk\bQ}, \theta_{\bp\bQ}) + \mathcal{A}^2(\theta_{\bk\bQ}, -\theta_{\bp\bQ})  \nn \\
= \frac{q^4 V_\bQ^2}{m^2} \left[ 4 \left( 1 -2 \cos^2 \theta_{\bq\bQ} \right)^2 + 
32 \frac{(\vf Q)^2}{\omega^2}\cos^2 \theta_{\bq\bQ} \sin^2 \theta_{\bq\bQ} \right]. \label{m}
\eea
On calculating the angular and energy integrals as  
$(2\pi)^{-1} \int^{2\pi}_0 d\theta_{\bq\bQ} \left( 1 -2 \cos^2 \theta_{\bq\bQ} \right)^2  = 1/2$ and 
$(2\pi)^{-1}\int^{2\pi}_0 d\theta_{\bq\bQ} \cos^2 \theta_{\bq\bQ} \sin^2 \theta_{\bq\bQ}  = 1/8$
and
\bea
\int^\infty_{-\infty} d\Omega  \int^\infty_{-\infty} d\ve_\bp \int^\infty_{-\infty} d \ve_\bk  n_\mathrm{F}(\ve_{\bp}) n_\mathrm{F}(\ve_{\bk}) \left(1-n_\mathrm{F}(\ve_{\bk} -\Omega)\right)\left(1-n_\mathrm{F}(\ve_{\bp}+\Omega+\omega)\right) =  \frac{\omega(\omega^2+ 4\pi^2T^2)}{6 (1- e^{-\omega/T})},  
\label{freqint}
\eea
respectively,
$\R\sigma$  is reduced to
\bea
\R\sigma(q,\omega)=\frac{e^2}{12\pi^2}\frac{q^2N_F^2}{\kf^2}\frac{\omega^2+4\pi^2 T^2}{\omega^2}\int \frac{dQ}{Q}V_\bQ^2 \left[1 + 2\left(\frac{\vf Q}{\omega}\right)^2 \right].\label{s_2}
\eea
With $V_\bQ$ given by Eq.~
\eqref{vsc}, the integrals over $Q$ in the last equation are calculated to leading logarithm accuracy as
    \bse
\bea
\int_{\mathrm{max}\{ |\omega|, T\}/\vf} \frac{dQ}{Q} \frac{1}{(Q+\kappa)^2} &\approx& \frac{1}{\kappa^2}\ln \frac{\vf \kappa}{\mathrm{max}\{|\omega|, T\}}\;\text{and}\label{logint} \\
\int^{k_F}_{0}\frac{dQQ}{(Q+\kappa)^2} &\approx&  \ln\frac{\kf}{\kappa}. \label{logint2}
\eea
\ese
Substituting the results of the $Q$-integration back into \Eq{s_2} gives \Eq{result2DEG} of the main text.

\subsection{3D electron gas}
\label{app:3DEG}
The starting point for the 3D case is an expression which differs from Eq.~\eqref{s22} only in the angular integrals:
\bea
\text{Re}\sigma(q,\omega) &=&\frac{e^2 (1- e^{-\omega/T})}{4(2\pi)^4\omega^3 q^2} N_F^2 \int_0^\infty dQ Q^2 \int^\infty_{-\infty}d \ve_\bp \int^\infty_{-\infty}d\ve_\bk  \int^\infty_{-\infty}d\Omega\int d\mathcal{O}_{\bQ\bq}\int d\mathcal{O}_{\bp\bQ}\int d\mathcal{O}_{\bk\bQ}\mathcal{A}^2 \nn \\
&\times& n_\mathrm{F}(\ve_\bp) n_\mathrm{F}(\ve_\bk) \left[1-n_\mathrm{F}(\ve_\bk -\Omega)\right]\left[1-n_\mathrm{F}(\ve_\bp+\Omega+\omega)\right] \delta\left(\bv_\bp\cdot\bQ
\right) \delta( \bv_\bk\cdot\bQ
), \nn \\ \label{3D}
\eea
where $d\mathcal{O}_{{\bf n}{\bf n}'}=dx_{{\bf n}{\bf n}'}d\phi_{{\bf n}}$,  $x_{{\bf n}{\bf n}'}=\cos\theta_{{\bf n}{\bf n}'}$,  $\theta_{{\bf n}{\bf n}}'$ is the polar angle of vector ${\bf n}$ measured with with respect to ${\bf n}'$, and $\phi_{{\bf n}}$ is the azimuthal angle of ${\bf n}$. Please note that, based on the result for the 2D case, we already discarded the subleading terms in the delta-functions.
Anticipating that, as in 2D, that Eq.~\eqref{A'} gives the leading contribution to the conductivity, we  
replace $\mathcal{A}$ by $\mathcal{A}_2$ in Eq.~\eqref{3D}   
 In spherical system, $\mathcal{A}$ can be written as 
\bea
\mathcal{A}
&=&\mathcal{A}_2= 
2V_\bQ 
\frac{q^2Q x_\bQ}{m^2\omega}\left[\kf(x_{\bp\bq}- x_{\bk\bq}) - Q x_{\bQ\bq}\right],
\label{A1_3D}
\eea
where $V_\bQ = 4 \pi e^2 /(Q^2+\kappa^2)$ is the screened Coulomb potential in 3D, $x_{\bp\bq} = x_{\bp\bQ} x_{\bQ\bq} + \sqrt{1- x_{\bp\bQ}^2 }\sqrt{1- x_{\bQ\bq}^2 } \cos(\varphi_\bp-\phi_\bQ)$ and similarly for $x_{\bk\bq}$.
Next, we take into account the constraints imposed by the delta-functions in Eq.~\eqref{3D}, i.e., $x_{\bp\bQ}=0$ and $x_{\bk\bQ}=0$ and also, using the condition $Q\ll \kf$, neglect the $Qx_{\bQ\bq}$ term in Eq.~\eqref{A1_3D}. Then Eq.~\eqref{A1_3D} is simplified to
\bea
\mathcal{A}=
2V_\bQ\frac{q^2}{m}\frac{\vf Q }{\omega}x_{\bQ\bq}\sqrt{1-x_{\bQ\bq}^2}\left[\cos(\phi_p-\phi_Q)-\cos(\phi_k-\phi_Q)\right].
\eea
Integrating over the angles, we obtain 
\bea
\int d\mathcal{O}_{\bQ\bq}
\int d\mathcal{O}_{\bp\bQ}\int d\,\mathcal{O}_{\bk\bQ}
\mathcal{A}^2\delta(\bv_\bp\cdot\bQ)\delta(\bv_\bk\cdot\bQ)=\frac{96\pi^3}{15} \frac{q^4V_Q^2}{m^2\omega^2}.
\eea
Using Eq.~\eqref{freqint} for the energy integration, we arrive at
\bea
R\sigma(q,\omega) &= & \frac{e^2 q^2 }{
60\pi\vf^2 m^2 \omega^2   } (4 \pi e^2)^2 N_F^2 \frac{\omega^2 + 4 \pi^2 T^2}{\omega^2}  
\int^\infty_0 \frac{dQ Q^2}{ (Q^2+\kappa^2)^2}
\nn \\
&= & \frac{e^2\kappa }{
720} \frac{q^2}{\kf^2}\frac{
\vf^2\kappa^2
}{
\omega^2 }  \frac{\omega^2 + 4 \pi^2 T^2}{\omega^2},
\eea
which is \Eq{result3DEG} of the main text.
\subsection{Doped monolayer graphene}\label{app:graphene}
 Expanding the graphene dispersion to $\mathcal{O}(q^2)$ as
\bea
\ve_{\bp+\bq} = 
v_\mathrm{D}|\bp+\bq|-\Ef =\ve_\bp +\bv_\bp\cdot\bq + \frac{q^2 \vd }{2p} \sin^2\theta_{\bp\bq}
\eea
with $\bv_\bp =\vd \bp/p$,
we obtain 
\bea
\frac{1}{\omega - \ve_{\bp+\bq} +\ve_\bp} &=&
\frac{1}{\omega} \left[1+ \frac{\bv_\bp\cdot\bq}{\omega}  +  \frac{q^2\vd}{2p}\sin^2\theta_{\bp\bq} +\left(\frac{\bq\cdot \bv_\bp}{\omega}\right)^2 \right], \\
\eea
and similarly for other terms  in Eq.~\eqref{generalA}. Substituting these expansions into Eq.~\eqref{generalA}, we obtain
\bea
\mathcal{A}&=& V_{\bk-\bk'}  \left[ \bv_\bp \cdot \bq  -\bv_{\bp'} \cdot \bq +\frac{q^2\vd}{2p}\sin^2\theta_{\bp\bq}+ \frac{q^2\vd}{2p'}\sin^2\theta_{\bp'\bq}+ \frac{(\bv_\bp\cdot \bq)^2 - (\bv_{\bp'} \cdot \bq)^2}{\omega} \right] \nn \\ 
&+&V_{\bp'-\bp} \left[  \bv_\bk \cdot \bq  -\bv_{\bk'} \cdot \bq + \frac{q^2\vd}{2k}\sin^2\theta_{\bk\bq}+ \frac{q^2v_\mathrm{D}}{2k'}\sin^2\theta_{\bk'\bq} + \frac{(\bv_\bk\cdot \bq)^2 - (\bv_{\bk'} \cdot \bq)^2}{\omega} \right].
\label{L00_app}
\eea
The momentum transfer $Q$ is defined by $\bp-\bp' = \bQ$, and from momentum conservation we have $ \bk'-\bk=\bQ+\bq$.
As for the parabolic case, the difference between $V_{\bk-\bk'}=V_{\bQ+\bq}$ and $V_{\bp-\bp'}=V_\bQ$ can be neglected.
Next, we split $\mathcal{A}$ into two parts as $\mathcal{A}=\mathcal{B}+\mathcal{C}$, where 
\bea
\mathcal{B}&=&
V_\bQ\bq\cdot\left(\bv_\bp+\bv_\bk-\bv_{\bp'}-\bv_{\bk'}\right)
=\varv_\mathrm{D}V_{\bQ}
\bq\cdot\left( \frac{\bp}{p}+ \frac{ \bk  }{k}-\frac{\bp-\bQ}{|\bp-\bQ|}  -\frac{\bk+\bQ+\bq}{|\bk+\bQ+\bq|}   \right).
\eea
Expanding $\mathcal{B}$ to $\mathcal{O}(q^2)$, we obtain
\bea
\mathcal{B}= \varv_\mathrm{D}V_{\bQ}  
\left[  \frac{\bq \cdot \bp}{p} +\frac{\bq \cdot \bk  }{k}-\frac{\bq\cdot(\bp-\bQ)}{|\bp-\bQ|} -\frac{\bq\cdot(\bk+\bQ)}{|\bk+\bQ|}  - \frac{q^2 \sin^2\theta_{\bk+\bQ,\bq}}{|\bk+\bQ|}  \right].  \label{L001} 
\eea
Note that if replace the magnitudes of the momenta in the $\mathcal{O}(q)$ part of the last equation (the first four terms), the resultant expression would vanish. To obtain a non-zero result for the $\mathcal{O}(q)$ part, we need to expand the magnitudes of the momenta near the Fermi surface. Performing such an expansion in the $\mathcal{O}(q)$ part  and replacing $|\bk+\bQ|$ by $\kf$ in  the last, $\mathcal{O}(q^2)$, term, we obtain
\bea
\mathcal{B}=V_\bQ\left\{\frac{q}{\kf}\left[(\ve_{\bp+\bQ}-\ve_\bp)\cos\theta_{\bq\bp}+(\ve_{\bk-\bQ}-\ve_\bk)\cos\theta_{\bq\bk}+\frac{Q}{\kf}(\ve_{\bp+\bQ}-\ve_{\bk-Q})\right]_1-\frac{\vd q^2}{\kf}\sin^2\theta_{\bk-\bQ,\bq}\right\}.
\eea
Finally, we recall that in our case of $Q\ll \kf$ the last term in $[\dots]_1$ can be neglected, while
$\theta_{\bk-\bQ,\bq}$ in the last term can be replaced $\theta_{\bk\bq}$. With these simplifications,
\bea
\mathcal{B}=V_\bQ\left\{\frac{q}{\kf}\left[(\ve_{\bp+\bQ}-\ve_\bp)\cos\theta_{\bq\bp}+(\ve_{\bk-\bQ}-\ve_\bk)\cos\theta_{\bq\bk}\right]-\frac{\vd q^2}{\kf}\sin^2\theta_{\bk\bq}\right\}.\label{B}
\eea
As  the remaining part of \Eq{L00_app}, $\mathcal{C}$, is already proportional to $q^2$, we can neglect $q$ everywhere else in that part. Then,
\bea
\mathcal{C} &=& V_\bQ \left\{ \frac{q^2\vd}{2}\left[\frac{\sin^2\theta_{\bp\bq}}{p}+ \frac{\sin^2\theta_{\bp-\bQ,\bq}}{|\bp-\bQ|}+\frac{\sin^2\theta_{\bk\bq}}{k} + \frac{\sin^2\theta_{\bk+\bQ,\bq}}{|\bk+\bQ|} \right]_2 \right. \nn \\
&+& \left. \frac{1}{\omega}\left[(\bv_\bp\cdot \bq)^2 - (\bv_{\bp-\bQ} \cdot \bq)^2 +(\bv_\bk\cdot \bq)^2 - (\bv_{\bk+\bQ} \cdot \bq)^2 \right]_3\right\}.\label{C}
\eea
Under the condition of $Q\ll \kf$, we can safely set $Q=0$ in the square brackets denoted by  $[\dots]_2$. Also, because $\mathcal{C}$ is already proportional to $q^2$, momenta $p$ and $k$ can be replaced by $\kf$. Then $[\dots]_2$ is reduced to
\bea
[\dots]_2=\frac{2}{\kf}\left(\sin^2\theta_{\bp\bq}+\sin^2\theta_{\bp\bk}\right).\label{sq2}
\eea 
The square brackets denoted by $\left[...\right]_3$ vanish at $Q=0$. Expanding 
to $\mathcal{O}(Q)$ and again replacing $p$ and $k$ by $\kf$, we obtain
as \bea
\left[...\right]_3 
=\frac{2q^2\vd^2Q}{\kf}\left(F_\bp-F_\bk\right),\label{sq3}
\eea
where \bea
F_{\bf n}=\cos\theta_{{\bf n}\bq}(\cos\theta_{\bQ\bq}-\cos\theta_{{\bf n}\bQ}).\label{Fn}\eea
Substituting Eqs.~\eqref{sq2} and \eqref{sq3} back to Eq.~\eqref{C}, we obtain
\bea
\mathcal{C}  = \frac{q^2\vd V_\bQ}{\kf}\left[\sin^2\theta_{\bp\bq}+ \sin^2\theta_{\bk\bq}+2\frac{\vd Q} {\omega} \left(   F_\bp   -F_\bk   \right)\right]. \label {L002}
\eea 
Adding up Eqs.~\eqref{B} and \eqref{L002}, we re-write $\mathcal{A}$ as the sum of the $\mathcal{}{O}(q)$ and $\mathcal{O}(q^2)$ parts:
\bse
\bea
\mathcal{A}&=&\mathcal{A}_1+\mathcal{A}_{2},\label{A1A2}\\
\mathcal{A}_1&=&\frac{qV_\bQ}{\kf} \left[- (\ve_\bp- \ve_{\bp-\bQ}) \cos\theta_{\bq\bp}  + (\ve_\bk- \ve_{\bk+\bQ}) \cos\theta_{\bq\bk}\right], 
\label{A1B}\\
\mathcal{A}_2&=&\frac{q^2\vd V_\bQ}{\kf}\left[\sin^2\theta_{\bp\bq}+2\frac{\vd Q} {\omega} \left(   F_\bp   -F_\bk   \right)\right], \label{A2B}
\label{39_app}
\eea
\ese
as given by \Eq{A_graphene} of the main text. 
On substituting $\mathcal{A}^2=\mathcal{A}_1^2+\mathcal{A}_2^2+2\mathcal{A}_1\mathcal{A}_2$ into Eq.~\eqref{sigma_wAgeneral} for the conductivity, we see that $\mathcal{A}_1^2$ and $\mathcal{A}_2^2$ give the $q$-independent and $\mathcal{O}(q^2)$ terms, respectively, while the angular integration nullifies the cross-term, $2\mathcal{A}_1\mathcal{A}_2$.
The $q$-independent part of $\R\sigma$ is exactly the same as calculated in Ref.~\cite{Sharma:2021} and thus needs not to be discussed here. In what follows, we focus on the $\mathcal{O}(q^2)$ part. 

First,
we assume  (and will confirm later) that typical momentum transfers satisfy $Q\gg|\omega|/\vd$. Then the first term in the square brackets in Eq.~\eqref{A2B} can be neglected compared to the first one, and $\mathcal{A}_2$ is reduced to 
\bea
\mathcal{A}_2  = \frac{2q^2\vd Q}{m^*\omega}V_\bQ  \left(   F_\bp   -F_\bk   \right),\label {L003}
\eea
where $m^*=\kf/\vd$.

 Substituting Eq.~\eqref{L003} into Eq.~\eqref{sigma_wAgeneral}, we obtain
 \bea
\R \sigma_2(\bq,\omega) &=&\frac{4 e^2 q^2  (1- e^{-\omega/T})}{(2\pi)^3   \omega^5 } N_v^2N_F^2\frac{\vd^2}{m^{*2}}\int d^2 Q Q^2 V^2_\bQ\int^\infty_{-\infty} d \ve_\bp  \int^\infty_{-\infty} d\ve_\bk    \int^\infty_{-\infty}  d\Omega 
 \int_{0}^{
 2
 \pi} d\theta_{\bp\bQ} \int_{0}^{
 2
 \pi} d\theta_{\bk\bQ}\nn\\
 &\times& n_\mathrm{F}(\ve_\bp) n_\mathrm{F}(\ve_\bk) \left[1-n_\mathrm{F}(\ve_\bk -\Omega)\right[\left[1-n_\mathrm{F}(\ve_\bp+\Omega+\omega)\right]
 \nn\\
&\times&
  \int_{0}^{
 2
 \pi} d\theta_{\bp\bQ} \int_{0}^{
 2
 \pi} d\theta_{\bk\bQ}\left(F_\bp-F_\bk\right)^2\delta(\bv_\bp\cdot\bQ) 
 \delta( \bv_\bk\cdot\bQ),
 \eea  
 where $N_v=2$ is the valley degeneracy. The two delta-functions in the last equation are the energy-conservation delta-functions in Eq.~\eqref{sigma_wAgeneral}, in which we neglected  frequencies $\omega$ and $\Omega$, and also expanded the dispersions to order $\mathcal{O}(Q)$. As before, these delta-functions impose the constraints $\cos\theta_{\bp\bQ}=\cos\theta_{\bk\bQ}=0$.
 Imposing these constraints and integrating over 
$\theta_{\bq\bQ}$, we get 
\bea
\R \sigma_2(\bq,\omega) &=&\frac{4e^2 q^2  (1- e^{-\omega/T})}{\omega^5 m^{*2}  (2\pi)^2}N_v^2 N_F^2  \int dQ Q V_Q^2 
\nn \\ 
&\times & \int^\infty_{-\infty} d \ve_\bp \int^\infty_{-\infty} d \ve_\bk  \int^\infty_{-\infty} d\Omega n_\mathrm{F}(\ve_\bp) n_\mathrm{F}(\ve_\bk) \left[1-n_\mathrm{F}(\ve_\bk -\Omega)\right]\left[1-n_\mathrm{F}(\ve_\bp+\Omega+\omega)\right].  \label{s2_g2}
\eea
 The energy integrals in \Eq{s2_g2} give
\bea
I &=& (1- e^{-\omega/T})  \int^\infty_{-\infty} d \ve_\bp \int^\infty_{-\infty} d \ve_\bk  \int^\infty_{-\infty} d\Omega\, n_\mathrm{F}(\ve_\bp) n_\mathrm{F}(\ve_\bk) \left[1-n_\mathrm{F}(\ve_\bk -\Omega)\right]\left[1-n_\mathrm{F}(\ve_\bp+\Omega+\omega)\right] \nn \\
&=& \frac{ \omega}{6}  (\omega^2 +4 \pi^2 T^2 ),
\eea
while the integral over $Q$ is already solved in \Eq{logint2}.
Combining everything together, we obtain
\bea
\R\sigma_2(\bq,\omega) &=&\frac{e^2}{24 \pi^2} \frac{q^2 \kappa^2}{m^{*2}\omega^2}\left(1 + 4 \pi^2\frac{ T^2}{\omega^2}\right) \ln \frac{k_F}{\kappa}, 
\eea  
with $\kappa = 8\pi N_F e^2$ in graphene, which is Eq.~\eqref{P} of the main text.

\section{Derivation of 
Eq.~\eq{sigma_Convolution}}
\label{app:convolution}
We consider Eq.~\eq{sigma_wAgeneral} for the case of
an electron gas 
with parabolic dispersion. In the limit of $\vf q/\omega\ll 1$, we can neglect $\bq$ in the 
Fermi functions, and write $\mathcal{A}= \mathcal{A}_1+\mathcal{A}_2$, where 
$\mathcal{A}_1$ and $\mathcal{A}_2$ are given in Eqs.~\eq{A1_2} and~\eq{A'}, respectively. 
On 
re-defining the integration variables as $\bp \to \bp + \bQ/2$, $\bk \to \bk - \bQ/2$ and $\Omega \to \Omega - \omega$, Eq.~\eqref{sigma_wAgeneral} becomes 
\bea
\text{Re}\sigma(q,\omega) &=& \frac{ 
e^2 (1- e^{-\omega/T})}{ (2\pi)^{3D-1}q^2\omega^3} \int d{^D}\!Q \int d^D\!p \int d^D\!k \int d\Omega\,  
\left\{ \frac{1}{m}\left[q^2V_\bQ+(\bQ\cdot\bq)(\bq\cdot\boldsymbol{\nabla}V_\bQ)\right] + \frac{2V_\bQ}{m^2\omega}(\bq\cdot\bQ)\left[\bq \cdot(\bp-\bk)\right] \right\}^2
\,
\nn\\
&&\times 
n_\mathrm{F}(\ve_{\bk-\bQ/2}) n_\mathrm{F}(\ve_{\bp+\bQ/2}) \left[1-n_\mathrm{F}(\ve_{\bp-\bQ/2})\right]\left[1-n_\mathrm{F}(\ve_{\bk+\bQ/2})\right]  
\delta(\bp\cdot \bQ/m-\Omega + \omega)\delta(\Omega - \bk\cdot\bQ/m).
\label{app_sigma_wAgeneral_1}
\eea 
Here we used that $\ve_{\bp+\bQ/2}-\ve_{\bp-\bQ/2}  = \bp\cdot \bQ/m$ and $\ve_{\bk-\bQ/2}-\ve_{\bk+\bQ/2} = -\bk\cdot\bQ/m$. Next, we re-write the 
$\bq\cdot(\bp-\bk)$ factor in the curly brackets in Eq.~\eqref{app_sigma_wAgeneral_1} as 
\bea \label{app_vector_decomp}
\bq\cdot(\bp - \bk) &=& (\bq\cdot {\hat \bQ}) \big[{\hat \bQ}\cdot(\bp-\bk)\big] - \bq\cdot\big\{\big[(\bp - \bk) \times {\hat \bQ} \big] \times {\hat \bQ}\big\}
\nn\\
&\to&
-(\bq\cdot {\hat \bQ}) \frac{m\omega}{Q} - \big[(\bp - \bk) \times {\hat \bQ} \big] \cdot ({\hat \bQ}\cdot\bq),
\eea
where $\hat\bQ=\bQ/Q$. 
At the last step, we used the fact that the $\delta$-functions in the last line of Eq.~\eq{app_sigma_wAgeneral_1} allow us to replace ${\hat \bQ}\cdot\bp \to m(\Omega - \omega)/Q$ and ${\hat \bQ}\cdot\bp \to m\Omega/Q$. 
Substituting Eq.~\eq{app_vector_decomp} back into Eq.~\eq{app_sigma_wAgeneral_1}, we write the conductivity as the sum $\text{Re}\sigma(q,\omega) = \text{Re}\sigma_a(q,\omega) + \text{Re}\sigma_b(q,\omega)$, where 
\bea
\text{Re}\sigma_a(q,\omega) &=& \frac{
e^2 (1- e^{-\omega/T})}{ (2\pi)^{3D-1}q^2\omega^3 m^2} \int d{^D}\!Q \int d^D\!p \int d^D\!k \int d\Omega\,  
\left[q^2V_\bQ+(\bQ\cdot\bq)(\bq\cdot\boldsymbol{\nabla}V_\bQ) - 2V_\bQ (\bq\cdot{\hat \bQ})^2\right]^2
\,
\nn\\
&&\times 
n_\mathrm{F}(\ve_{\bk-\bQ/2}) n_\mathrm{F}(\ve_{\bp+\bQ/2}) \left[1-n_\mathrm{F}(\ve_{\bp-\bQ/2})\right]\left[1-n_\mathrm{F}(\ve_{\bk+\bQ/2})\right]  
\delta(\bp\cdot \bQ/m-\Omega + \omega)\delta(\Omega - \bk\cdot\bQ/m),
\label{app_sigma_wAgeneral_L}
\eea 
and
\bea
\text{Re}\sigma_b(q,\omega) &=& \frac{
4e^2 (1- e^{-\omega/T})}{ (2\pi)^{3D-1}q^2\omega^5 m^4} \int d{^D}\!Q \int d^D\!p \int d^D\!k \int d\Omega\,  
V_\bQ^2 (\bq\cdot\bQ)^2
\big\{ \big[(\bp - \bk) \times {\hat \bQ} \big] \cdot ({\hat \bQ} \times \bq) \big\}^2 
\,
\nn\\
&&\times 
n_\mathrm{F}(\ve_{\bk-\bQ/2}) n_\mathrm{F}(\ve_{\bp+\bQ/2}) \left[1-n_\mathrm{F}(\ve_{\bp-\bQ/2})\right]\left[1-n_\mathrm{F}(\ve_{\bk+\bQ/2})\right]  
\delta(\bp\cdot \bQ/m-\Omega + \omega)\delta(\Omega - \bk\cdot\bQ/m).
\label{app_sigma_wAgeneral_T}
\eea 
We have neglected the products of the two terms in Eq.~\eq{app_vector_decomp}, since they vanish on angular integration for a homogeneous electron gas. Indeed, that delta-functions in Eq.~\eq{app_sigma_wAgeneral_1} fix the value of the cosine of the angles between $\bp$ and $\bQ$, and between $\bk$ and $\bQ$. For the products of the two terms in Eq.~\eq{app_vector_decomp}, the solutions of the delta functions yield contributions which are equal in magnitudes but opposite in signs. This is because the second term in Eq.~\eq{app_vector_decomp} depends on the sine of the angles between $\bp$ and $\bQ$, and $\bk$ and $\bQ$.

We first focus on Eq.~\eq{app_sigma_wAgeneral_L}, which can be further simplified by assuming that $V_\bQ \propto (Q^{D-1}+\kappa^{D-1})^{-1}$. Then,
\bea
\text{Re}\sigma_a(q,\omega) &=& \frac{
(1- e^{-\omega/T})}{ (2\pi)^{3D-1}q^2\omega^3 m^2} \int d{^D}\!Q d^D\!p d^D\!k d\Omega\,  V_\bQ^2 
\left[q^2 - \frac{(D-1)Q^{D-1}}{Q^{D-1}+\kappa^{D-1}} ({\hat \bQ}\cdot\bq)^2 - 2 (\bq\cdot{\hat \bQ})^2 \right]^2
\,
\nn\\
&&\times 
n(\ve_{\bk-\bQ/2}) n(\ve_{\bp+\bQ/2}) \left[1-n(\ve_{\bp-\bQ/2})\right]\left[1-n(\ve_{\bk+\bQ/2})\right]  
\delta(\bp\cdot \bQ/m-\Omega + \omega)\delta(\Omega - \bk\cdot\bQ/m).
\label{app_sigma_wAgeneral_L_2}
\eea
We note that, using the $\delta$-functions, one can rewrite
\bea \label{app_bose_fermi}
(1- e^{-\omega/T}) n(\ve_{\bk-\bQ/2}) n(\ve_{\bp+\bQ/2}) \left[1-n(\ve_{\bp-\bQ/2})\right]\left[1-n(\ve_{\bk+\bQ/2})\right]  &\to&
\big[ n_B(\Omega) - n_B(\Omega - \omega) \big]
\big[n(\ve_{\bk-\bQ/2}) - n(\ve_{\bk+\bQ/2}\big]
\nn\\
&& \times
\big[n(\ve_{\bp+\bQ/2}) - n(\ve_{\bp-\bQ/2}) \big],
\eea
where $n_B(\omega) = (e^{\omega/T} - 1)^{-1}$ is the Bose distribution function.
Using this expression and introducing the density-density response function as in Eq.~\eq{spectrum_density_fluctuations}, we re-write Eq.~\eq{app_sigma_wAgeneral_L_2} 
as
\bea
\text{Re}\sigma_a(q,\omega) &=& \frac{1}{2 q^2\omega^3 m^2} \int \frac{d{^D}\!Q}{(2\pi)^D} \int_{-\infty}^\infty \frac{d\Omega}{\pi}\,  V_\bQ^2 
\left[q^2 - \frac{(D-1)Q^{D-1}}{Q^{D-1}+\kappa^{D-1}} ({\hat \bQ}\cdot\bq)^2 - 2 (\bq\cdot{\hat \bQ})^2 \right]^2 \big[ n_B(\Omega) - n_B(\Omega - \omega) \big]
{\text Im}\chi_c(Q,\Omega) 
\nn\\
&&\times
{\text Im}\chi_c(Q, \Omega-\omega)
\nn\\
&=&
\frac{q^2}{\omega^3 m^2} \int_0^\infty \frac{d^D\!Q}{(2\pi)^{D}} \int_{-\infty}^\infty \frac{d\Omega}{\pi}\,  V_\bQ^2 a_D(Q)
\big[ n_B(\Omega) - n_B(\Omega - \omega) \big]
{\text Im}\chi_c(Q,\Omega) 
{\text Im}\chi_c(Q, \Omega-\omega),
\label{app_sigma_wAgeneral_L_3}
\eea
where
\bea
&& a_3(Q) = \frac{23 Q^4 + 18 Q^2 \kappa^2 + 7 \kappa^4}{30 (Q^2 + \kappa^2)^2} \stackrel{\kappa \to 0}{\longrightarrow} \frac{23}{30},
\nonumber\\
&& a_2(Q) = \frac{11 Q^2 + 12 Q \kappa + 4 \kappa^2}{16 (Q + \kappa)^2} \stackrel{\kappa \to 0}{\longrightarrow} \frac{11}{16}.
\eea
Note that we are allowed to take the limit of $\kappa \to 0$ in $a_D(Q)$ in the limit of weak interaction since 
the integral in Eq.~\eq{app_sigma_wAgeneral_L_3} remains convergent in this limit.

We now focus on Eq.~\eq{app_sigma_wAgeneral_T}. Introducing the density-density response function 
as per Eq.~\eq{spectrum_density_fluctuations} and the current-current response function
\begin{equation} \label{spectrum_current_fluctuations}
\I \chi_{\alpha\beta}(\bQ,\nu) \equiv -2\pi \int \frac{d^D\!k}{(2\pi)^D} (n_{{\bf k}+{\bf Q}/2} - n_{{\bf k}-{\bf Q}/2}) \frac{k_\alpha k_\beta}{m^2} \delta({\bf k}\cdot{\bf Q}/m
-\nu),
\end{equation}
we obtain for Eq.~\eq{app_sigma_wAgeneral_T}
\bea
\text{Re}\sigma_b(q,\omega) &=& \frac{4 e^2}{q^2\omega^5 m^2} \int \frac{d{^D}\!Q}{(2\pi)^D} \int_{-\infty}^\infty \frac{d\Omega}{\pi}\,  
V_\bQ^2 (\bq\cdot\bQ)^2
\epsilon_{\alpha\beta\gamma} \epsilon_{\alpha'\beta'\gamma'} {\hat \bQ}_\beta {\hat \bQ}_{\beta'}  ({\hat \bQ} \times \bq)_{\gamma}  ({\hat \bQ} \times \bq)_{\gamma'}
\nn\\
&& \times
\big[ n_\mathrm{B}(\Omega) - n_\mathrm{B}(\Omega - \omega) \big] 
\I\chi_{\alpha\alpha'}({\bm Q},\Omega - \omega)
\I\chi_c (Q,\Omega) 
,
\label{app_sigma_wAgeneral_T_2}
\eea
where $\epsilon_{\alpha\beta\gamma}$ is the Levi-Civita tensor and sum over repeated Greek indices is implied. 
Owing to the isotropy of the electron gas, the current-current response function can be written as
\beq \label{spectrum_current_decomp}
\chi_{\alpha\beta}(\bQ,\nu) = \frac{Q_\alpha Q_\beta}{Q^2} \chi_{L}(Q,\nu) + \left( \delta_{\alpha\beta} - \frac{Q_\alpha Q_\beta}{Q^2}\right) \chi_{T}(Q,\nu),
\eeq
where 
$\chi_{L}(Q,\nu)$ and $\chi_{T}(Q,\nu)$ are the longitudinal and transverse current-current response functions, respectively. The imaginary part of the latter is given in Eq.~\eq{spectrum_transverse_fluctuations}, while the former is connected to $\chi_c (Q,\nu)$ by the relation $\nu^2 \chi_c (Q,\nu) = Q^2 \chi_{L}(Q,\nu)$. Note that 
both $\chi_L$ and $\chi_T$
depend only on the 
magnitude
of $\bQ$, a fact that reflects the isotropy and rotational invariance of the electron gas. It  
can be 
readily seen that only the term proportional to the Kronecker delta in Eq.~\eq{spectrum_current_decomp} contributes to Eq.~\eq{app_sigma_wAgeneral_T_2}, which thus becomes
\beq
\text{Re}\sigma_b(q,\omega) = \frac{4 e^2}{ q^2\omega^5 m^2} \int \frac{d{^D}\!Q}{(2\pi)^D} \int_{-\infty}^\infty \frac{d\Omega}{\pi}\,  
V_\bQ^2 (\bq\cdot\bQ)^2 |{\hat \bQ} \times \bq|^2
\big[ n_\mathrm{B}(\Omega) - n_\mathrm{B}(\Omega - \omega) \big] 
\I\chi_T(Q,\Omega - \omega)
\I\chi_c (Q,\Omega),
\label{app_sigma_wAgeneral_T_3}
\eeq
Finally, performing the angular integration we get
\beq
\text{Re}\sigma_b(q,\omega) =  b_D \frac{e^2 q^2}{\omega^3 m^2} \int \frac{d{^D}\!Q}{(2\pi)^D} \int_{-\infty}^\infty \frac{d\Omega}{\pi}\,  
\frac{Q^{2}}{\omega^2} V_\bQ^2
\big[ n_\mathrm{B}(\Omega) - n_\mathrm{B}(\Omega - \omega) \big] {\text Im} \chi_T(Q,\Omega - \omega)
{\text Im} \chi_c (Q,\Omega),
\label{app_sigma_wAgeneral_T_4}
\eeq
where $b_D = 8/15$ for $D = 3$ and $b_D = 1/2$ for $D=2$. We note that the results given in Eqs.~\eq{app_sigma_wAgeneral_L_3} and~\eq{app_sigma_wAgeneral_T_4}  agree with those of Ref.~\cite{nifosi:1998} once the conductivity is converted into the dynamical exchange-correlation potential $f_{\rm xc}$ as~\cite{giuliani,nifosi:1998}
\beq
\text{Im }f_{\rm xc}(\omega) = \lim_{q\to 0} \frac{m^2 \omega^3}{n^2 e^2 q^2} \text{Re}\sigma(q,\omega),
\eeq
where $n$ is the carrier number density.

\ewt

\end{document}